# Causality and Relativity in Quantum Physics

**Johan F. Prins**

Department of Physics, University of Pretoria, Gauteng, South Africa

**Postal address directly to author:** P O Box 1537, Cresta 2118, Gauteng, South Africa

**Telephone:** +27 11 477-8005

**Facsimile:** +27 11 477-3709

e-mail:  johanprins@cathodixx.com

**Abstract**

It is argued here that the Copenhagen interpretation of quantum mechanics violates the tenets on which both Galileo's and Einstein's theories of relativity are based. It is postulated that the "building blocks" of the universe are not particles but are holistic wave-entities which act and interact with other wave-entities as one would expect from waves: i.e. they can change their shape and spatial extent (morph) when the boundary conditions change, and they can superpose when sharing the same region of space. It is concluded that there are two distinctly different ways in which superposition can occur: (i) the holistic waves can add without totally losing their separate identities so that the superposed wave can experience interactions between its sub-components (it is suggested that this process should be termed "enmeshment"); (ii) the holistic waves can add to completely lose their separate holistic identities in order to form a wave which is again a single holistic entity (it is suggested that this process be termed "entanglement"). A parameter is derived which defines an interface between quantum mechanics and classical mechanics: the one does not encompass the other; both are valid within their respective domains. Using this framework, a possible causal interpretation is developed for quantum mechanics which can be visualised; even the "spooky action at a distance" that worried Einstein so much, seems to follow logically. This analysis suggests a new approach to model the electron (and other fundamental particles) quantum-mechanically. The electromagnetic quantum-energy of a solitary electron does not form a field around the electron in three-dimensional space, but is finite and localised so that it is equal to the mass of the electron. An energy component along the fourth dimension is derived for the electron which could be part of the dark energy of the Universe and which, in turn, causes vacuum-energy fluctuations through Heisenberg's uncertainty relationship for energy and time. The possible consequences of this approach are analysed and discussed.

**Contents**

1. Introduction

*General relativity explains big things, quantum mechanics explains little things, and if the twain meets it is on a scale that is physically undetectable and hence empirically irrelevant.*
                                                        *Freeman Dyson as quoted in New Scientist*

*The unpredictable, random element comes in only when we try to interpret the wave in terms of the positions and velocities of particles. But maybe that is our mistake: maybe there are no particle positions and velocities of particles, but only waves.*
                                                        **Stephen Hawking in "A Brief History of Time"**

Notwithstanding opposition from formidable scientists like Einstein, Schrödinger, de Broglie and Bohm, the Copenhagen interpretation of quantum mechanics still reigns supreme after its inception 80 years ago. It is at present generally believed that realism in science has been defeated; the quantum world does not exist unless one looks for it, and in this process one becomes a creator of the "reality" one experiences. This interpretation rests primarily on two principles postulated by Born and Heisenberg, and further "rationalised" by Bohr when he formulated his principle of complementarity according to which the particle and wave behaviour are two aspects of the same "reality" which cannot manifest simultaneously:

(i) According to Born the intensity of the wave-function, derived from the Schrödinger equation, is not the intensity of a "real matter wave" but a probability distribution representing the spread in possible positions at which a quantum particle (like an electron) will be found when a measurement is made.

(ii) According to Heisenberg there are inbuilt uncertainties in nature. When, for example, measuring position and momentum, uncertainties must appear even if perfect measurements were possible; i.e. in terms of Planck's reduced constant there exists along any direction in space a relationship between the uncertainty $\Delta r$ in position and $\Delta p$ in momentum of a particle; namely

$$\Delta r \Delta p \geq \tfrac{1}{2} \hbar \qquad (1)$$

This relationship implies that the product of $\Delta r$ with $\Delta p$ can at best be equal to $\tfrac{1}{2}\hbar$. In terms of the Born-interpretation, it implies that an accurate measurement of the position of a particle along the direction r, must lead to an infinitely large uncertainty in the momentum and *vice-versa*.

These two principles logically lead to the conclusion that the position and momentum of a quantum particle cannot manifest in such a way that both can be measured or known simultaneously, and that repeated measurements of any one of these parameters under exactly the same physical conditions give different results for each measurement; i.e. "God plays dice!". The Copenhagen interpretation has become entrenched in the scientific world. In fact, it has become dangerous to advocate an objective reality for the quantum world. It is nowadays widely believed that if one attempts to do so, it implies that one does not "understand" quantum physics.

In 1935 Einstein took a "last stand" against the Copenhagen interpretation when he, Podolsky and Rosen formulated their (EPR) paradox [1]. The experiment is based on the concept of "entanglement" of particles. Schrödinger coined the term "entanglement" when particles are represented by a single multi-particle wave-function. It implies an instantaneous "understanding" between the particles when they are modelled by quantum mechanics; a "spooky action at a distance". Einstein *et al* pointed out that this implies that two "entangled" particles separated by light years must still be able to communicate instantaneously with each other when a measurement is made on one of them. The other particle must then "immediately" manifest an outcome that is determined by the outcome of the measurement on its partner light years away. Einstein *et al* reasoned that (according to Einstein's theory of relativity) faster-than-light communication is not possible between two such particles, and therefore the instantaneous communication required by entanglement must indicate that the theory of quantum mechanics is incomplete. In 1964 John Bell derived a condition that has to be violated if such instantaneous communication can manifest [2]. Subsequent experiments on entangled photons violated Bell's inequality [3]. This has led to the conclusion that "Einstein has been proved wrong"; our universe is indeed the bizarre, non-causal, statistical concoction portrayed by the Copenhagen interpretation.

In this publication it will be argued that the Copenhagen interpretation violates fundamental relativistic concepts (from Galileo to Einstein), and that this has led to the present conundrums in physics; i.e. when a quantum "particle" with mass is analysed within its correct relativistic context it is found that a causal interpretation for quantum mechanics **must** apply.

**2. Relativity**

When talking about relativity Einstein's name immediately comes to mind; however, the real father of relativity is Galileo. He was first to consider the implications when bodies move relative to one another. He concluded that there is no mechanical experiment that one can do to determine whether you (the observer) and the "stationary surroundings" you observe around you are moving or not moving with a constant speed. From this he postulated that the earth is moving and nearly lost his life for parting with accepted scientific dogma.

Galileo's conclusions about relativity later became formalised in Newton's first law of classical mechanics. For the purposes of this publication, Newton's first law will be restated as follows: a body with mass that moves with a constant velocity is a "stationary body"; this is so because one can attach a reference frame to such a body (its proper reference frame) within which it will be observed as being stationary and within which, according to Galileo, it will act like a "real" stationary body. The body's mass can then be interpreted as its "inertia"; i.e. its "resistance against being moved" from its stationary position within its proper reference frame. It is well-known that it is for this reason that reference frames which move with constant velocities relative to each other have become known as "inertial" reference frames. The fact that a body with mass which moves with a constant velocity can be considered as stationary, does not exclude the possibility that there might be a unique stationary reference frame in the universe relative to which all other inertial reference frames move. For years it has been believed that the ether, within which light waves supposedly had to manifest and which on its own must thus be a static medium that fills the whole universe, can serve as such a reference frame; however, the failure of the Michelson-Morley experiment to detect the movement of the earth relative to the ether, and the derivation by Maxwell of his famous electromagnetic wave equations led Einstein to postulate that such a special inertial reference frame does not exist.

In his special theory of relativity Einstein amended Galilean relativity by postulating that all inertial reference frames are equivalent, and that there is no experiment possible, even when using light, which an observer can use to prove that his/her stationary surroundings are moving or not! This forced him to further postulate that the speed of light must be the same within whichever inertial reference frame it is measured. It is accepted in the scientific literature today that light consists of photons ("particles"), and that a photon does not have mass. According to the discussion above, this is exactly as it has to be. If a photon had mass, it would have been able to resist being moved from a stationary position within an inertial reference frame; i.e. a photon can therefore not have a proper inertial reference frame within which it can be at rest. As is well known, two major consequences of these postulates by Einstein are (i) that time relates to a fourth dimension of the universe and (ii) that mass is the same as energy; i.e.

$$E = mc^2 \qquad (2)$$

## 3. Time-independent wave-functions of an electron

It is well known that Schrödinger's time-dependent equation for an electron can be written as:

$$H\psi(\mathbf{r},t) = -\frac{\hbar^2}{2m_e}\nabla^2\psi(\mathbf{r},t) + V(\mathbf{r},t)\psi(\mathbf{r},t) = i\hbar\frac{\partial}{\partial t}\psi(\mathbf{r},t) \qquad (3)$$

The symbols have their usual meaning. H is the Hamilton operator which is interpreted as the sum of the kinetic energy and potential energy in operator format; i.e.

$$H = \frac{p^2}{2m_e} + V(\mathbf{r},t) \qquad (4)$$

In other words the momentum operator can be written as $\mathbf{p} = -i\nabla$. When the potential energy $V(\mathbf{r},t)$ of the electron does not change with time, so that $V(\mathbf{r},t)=V(\mathbf{r})$, the solution of the equation renders a complete basis-set of wave-functions where each has a definite energy $E_n$ enumerated by a quantum number or a set of quantum numbers. The quantum numbers are determined by the boundary conditions defined by $V(\mathbf{r})$. In general, when these wave-functions are localised, in the sense that they primarily occupy a small region in space, the differences between consecutive energies are large; when the wave-functions spread over large regions in space, these differences become smaller. Each one of these wave-functions can be written in terms of its concomitant energy $E_n$ as:

$$\psi_n(\mathbf{r},t) = \psi_n(\mathbf{r})\exp\left[-i\frac{E_n}{\hbar}t\right] = \psi_n(\mathbf{r})e^{-i\omega_n t} \qquad (5)$$

When calculating the density of the wave intensity $\rho_n(\mathbf{r})$ in the normal manner, one obtains that:

$$\rho_n(\mathbf{r}) = \psi_n(\mathbf{r},t)^* \psi_n(\mathbf{r},t) = |\psi_n(\mathbf{r})|^2 \qquad (6)$$

The total intensity $I_n$ of the wave can then be calculated by integrating over the region of space which the wave occupies; i.e.

$$I_n = \int_{\text{volume}} \rho_n(\mathbf{r})\,d\tau \qquad (7)$$

where $d\tau$ is an infinitesimal volume at the point denoted by the position vector $\mathbf{r}$. If, as postulated by Born, the wave-intensity represents a probability distribution, then $I_n$ is the probability of finding the electron anywhere "within the wave" and therefore $I_n$ must be equal to unity. The resultant wave-intensity of any one of the basis-solutions is completely time-independent provided that the potential energy does not change in any way. It must be emphasised that such a stationary wave does not imply the same state as, for example, a stationary wave on a violin string. The latter wave is a superposition of waves that move in opposite directions along the string. In this case the resultant superposed wave still changes with time; i.e. the string still vibrates. In the case of a single-electron stationary-wave there is no change whatsoever at any point in space with time. Thus, for example, the intensity of the wave representing an electron-orbital around the nucleus of an atom does not oscillate with time like a standing wave on a string. But if the wave-intensity is unchangingly stationary in space, what is it stationary relative to? Must it not be stationary relative to a proper inertial reference frame? Relative to another inertial reference frame the position vectors $\mathbf{r}$ referencing the points within the wave will have to change with time. The wave will then be observed as a moving wave; i.e. it will change with time!

Each one of the stationary basis wave-functions can represent an electron which, owing to its postulated "probabilistic wave-nature", is interpreted to have uncertainties in its position and momentum. As already pointed out above (according to the Copenhagen interpretation) the uncertainty-spread in position gives the spectrum of possible positions that are able to manifest when a position measurement is made, while the uncertainty-spread in momentum gives the spectrum of possible momentums that can manifest when, instead, the momentum of the same electron is measured. But if this is the case, does it not imply that the electron constituting an atomic orbital must be "moving around within" such an orbital relative to the positive charge on the nucleus. If it is not "moving around" one should find it exactly at the same position and with zero momentum every time when making the same position measurement. The only "logical" argument (consistent with the Copenhagen interpretation) to prevent this conclusion is to postulate that the electron is not really there until one does the measurement! But what is then there? Electrons are, for example, responsible for forming covalent bonds and must thus be "there" for the bond to keep on manifesting even when one is not making a measurement to determine the position of any one of the two electrons. Thus, when applying Heisenberg's uncertainty relationship (as interpreted by Heisenberg) to such a bond, one again has to conclude that the two electrons are moving about. But when charges move relative to each other, Maxwell's equations demand that electromagnetic waves should be radiating. So why does this not happen?

It is now proposed here that the latter problem should be removed by postulating that the electron **is** the time-independent wave intensity of the orbital; i.e. the electron is **not** a particle with a point charge and an uncertainty in its position and momentum, but the electron is in this case a "localised" field within which its charge is "smeared out" within the wave intensity; i.e. the wave-field is a holistic single-entity which is distributed over a region of space. "Localised" in this sense implies that the stationary, "smeared-out" charge-intensity has a centre of charge around which the charge is distributed within a relatively small volume. If this stationary-distributed charge is the reality instead of a particle with a point charge, then its concomitant charge distribution is completely stationary relative to the other charges which are sharing a proper inertial reference frame with it; e.g. the positive charge of the nucleus. Therefore electromagnetic waves cannot radiate. This viewpoint also implies that "the electron" must be in "immediate contact" with itself over the space-volume it occupies; i.e. time cannot exist "within" such a holistic single wave-entity.

But what about a "free electron" which is moving through space far away from any other charges? Can this entity be interpreted as a localised wave with a centre of charge instead of being a point-charge particle?

**4. Traditional modelling of a "free" electron**

One might want to argue that it is a waste of time to model a free electron in terms of Schrödinger's "non-relativistic" equation. It would be better to use Dirac's equation. For reasons that will become clear we will start off with Schrödinger, and comment later on Dirac's equation (see section 14.2).

In all textbooks the basis-solutions of the Schrödinger equation for a free electron gives "extended" solutions; i.e. the wave-intensity of such a wave is not localised. These wave-functions are obtained from the Schrödinger equation by setting the potential energy V(**r**,t) of the electron identically equal to zero; i.e. the potential energy of the electron does not change with time or position. It is then found that the resultant basis functions are given by:

$$\psi_n(\mathbf{r},t) = C_n \exp[i(\mathbf{k}_n \bullet \mathbf{r} - \omega_n t)] \qquad (8)$$

The different values of $\mathbf{k}_n$ are the appropriate wave vectors of these complex wave-functions. The probability distribution for such a wave follows as:

$$\rho_n(\mathbf{r},t) = \psi_n(\mathbf{r},t)^* \psi_n(\mathbf{r},t) = |C_n|^2 \qquad (9)$$

When normalising the intensity by integration it is found that all the basis functions have the same constant intensity over all space, and that each one of these "probability" distributions is stationary over the whole universe. The question to ask is: stationary relative to which inertial reference frame? Do these functions not require a unique, universal, stationary inertial reference frame? I believe that they do, because the electron is modelled as if there is no other matter present in the universe. Such a reference frame is not possible according to Einstein's theories of relativity. So is it reasonable to accept that these solutions can manifest? Furthermore, in terms of the statistical interpretation of the wave-function (as postulated by Born) it will be near impossible, if not totally impossible, to find a "free" electron in the universe. Experimentally this is not the case. Experimental evidence must reign supreme. So what is wrong?

The traditional method used in textbooks to "make this problem go away", is not to normalise the intensities of the wave-functions but to superpose them in order to form a localised wave "packet"; which is then normalised. This can be done by summing over all the allowed wave-functions; the localised wave is then generated by having a suitable distribution in the values of $C_n$ for different values of $\mathbf{k}_n$. Since the basis waves are spread over the whole universe, the increments in energy (and thus the wave vectors) can be taken as near-continuous so that a wave packet can be written as a Fourier-integral: i.e.

$$\psi(\mathbf{r},t) = \frac{1}{(2\pi)^{3/2}} \int_{-\infty}^{+\infty} \phi(\mathbf{k},t) e^{i[\mathbf{k} \bullet \mathbf{r} - \omega(\mathbf{k})t]} d\tau_k \qquad (10)$$

The $C_n$'s have been replaced with the continuous function $\phi(\mathbf{k},t)$ in wave-vector space. Thus knowing this function, one can derive the wave-function $\psi(\mathbf{r},t)$. Alternatively when knowing the wave-function one can derive the concomitant function in **k**-space from the following, inverse Fourier-integral:

$$\phi(\mathbf{k},t) = \frac{1}{(2\pi)^{3/2}} \int_{-\infty}^{+\infty} \psi(\mathbf{r},t) e^{-i[\mathbf{k}\cdot\mathbf{r}-\omega(\mathbf{k})t]} d\tau \qquad (11)$$

$\phi(\mathbf{k},t)$ is also a "wave packet" which, however, manifests in wave vector space (**k**-space). At any time t, say t=0, one can calculate the spread in position $\Delta r$ of the intensity of the wave packet along a chosen direction, and also the concomitant spread in the wave vector $\Delta k$ along the same direction using the intensity of the wave packet in **k**-space; i.e. this is done by making use of the functions $\psi(\mathbf{r},0)$ and $\phi(\mathbf{k},0)$ respectively. Owing to the inverse relationship between the two Fourier-integrals it is found that the product $\Delta r \Delta k$ can never be zero. The smallest number (they can be equal to) is found when $\psi(\mathbf{r},0)$ has a Gaussian shape along the direction r. It is then found that $\phi(\mathbf{k},0)$ also has a Gaussian shape along the same direction. In this case one has that:

$$\Delta r \Delta k = \frac{1}{2} \qquad (12)$$

Thus if the average spread of the wave intensity along a direction in position space is large, then the average spread along the same direction of the wave intensity in **k**-space must be small and *vice versa*. This is reminiscent of Heisenberg's uncertainty relationship (see Eq. (1)). In fact, it is reasoned in textbooks that this forms the basis for the Heisenberg relationship and the statistical interpretation of the wave-function. According to de Broglie's hypothesis, the momentum of an electron is proportional to the wave vector: i.e.

$$\mathbf{p} = \hbar \mathbf{k} \qquad (13)$$

Thus by writing for $\Delta p = \hbar \Delta k$, Eq. (13) becomes Eq. (1). The following problem, however, still remains: relative to which inertial reference frame is such a wave packet described? The solution still requires a uniquely stationary reference frame for the basis functions and thus also for the wave packet; and this does not exist. Furthermore, the Schrödinger equation describes "an electron". So each possible solution must represent "an electron". Thus a wave packet can only be generated by superposing waves, each of which represents an electron. It is known (and fits experiment) that it is possible within a conducting solid material to superpose extended (so-called delocalised) valence-electron waves in order to form a wave packet. Although the resultant wave-packet (or excitation) can under suitable circumstances act very much like a single electron, it is not one; therefore it is referred to as a pseudo-particle. Does it then make sense to construct an actual electron in free space in the same manner? It seems unlikely. In addition, when constructing such a wave packet, the exponential, energy-time factors (see Eq. (10)) have to be included. It is then found that such a wave packet, when left to its own devices, spreads out in space with time; i.e. the uncertainty in the position of the electron increases with time. Has this ever been observed experimentally for an electron in free space? Not that I know off! So why do scientists believe this model?

**5. The "path" of an electron in space**

The accepted quantum-mechanical model for a free electron in space (discussed in section 4) is at variance with what is known experimentally about electrons. Electrons were discovered as cathode rays and they do not become "uncertain" once they are ejected from a cathode. In fact they follow well defined paths which can be accurately modelled by means of classical mechanics. It is this behaviour of electrons that convinced Born that "particles" are the reality and that the wave-function's intensity should be interpreted as a probability distribution; however, to model the path classically, both the position and momentum of the centre of mass of an electron must be specified at each instance in time. This is not possible according to Heisenberg's interpretation of his uncertainty relationship. Heisenberg addressed this problem by the following statement: *"I believe that the existence of the classical "path" can be pregnantly formulated as follows: The "path" comes into existence only when we observe it"*. This implies that the quantum world "magically" becomes the classical world when an observer is present. "Suddenly" it is possible to violate Heisenberg's own interpretation of his uncertainty relationship for position and momentum!

How can one observe the path if it is not possible to simultaneously observe the position and momentum of the electron? One possibility (and probably the only one) would be to conclude that the position and momentum manifest sequentially: i.e. one first observes the position, then the momentum, then again the position, etc. But when, for example, observing the position of the particle, then, according to the Copenhagen interpretation, the momentum becomes indeterminate; when next observing the momentum, any momentum from a spectrum of possible momentums might manifest. Similarly when next again observing the position, any position out of a spectrum of possible positions might manifest. In other words, the kinetic energy of the electron will increase and decrease erratically and its position will also jump around erratically; i.e. the "path" of the electron should be akin to Brownian motion. There is not a single experimental result that shows such behaviour for an electron when it moves through free space. If this were the reality, there would not be operational electron microscopes or electron accelerators. Furthermore, such random jumps in energy violate the conservation of energy; where does the energy come from or go to each time when the momentum changes statistically?

One must always be led by experimental results. As already mentioned above, there exists no doubt that the path of an electron (or at least its centre of mass) in free space can be accurately modelled by specifying both the position of its centre of mass and its momentum at the same time. These calculations have been accurately consistent with all experiments done to date. As discussed in section 2, a particle with mass that moves with a constant velocity must be stationary within its proper inertial reference frame moving with it. A free electron with no forces on it can be accurately modelled as a particle moving through space with a constant velocity; in fact, it must be so because the electron has mass. As discussed in section 2, mass is the resistance of a particle against being moved from its stationary position within its proper inertial reference frame. Since the electron must be stationary within its proper inertial reference frame and since such a reference frame moves relative to other inertial reference frames, it **must** always be possible to simultaneously specify both the position and momentum of the centre of mass of a free electron relative to any inertial reference frame. If this reasoning is correct, it means that the Copenhagen interpretation **must** be wrong. This, in turn, means that Heisenberg's uncertainty relationship cannot have anything to do with uncertainties in the position and momentum of a particle. It requires another explanation.

**6. Heisenberg's "uncertainty" relationship**

A photon's wave vector **k** always corresponds to the photon's momentum through the de Broglie relationship (Eq. (13)). Therefore, if there is an uncertainty in its wave vector, this must correspond to an uncertainty in its momentum. But the photon always moves with the same speed relative to all inertial reference frames. This is not the case for a particle with mass. Consider a time-independent, localised matter-wave within its proper reference frame. Does the uncertainty in the wave vector of the time-independent stationary field relate to momentum in this case? I cannot see how this is possible, because the momentum only comes into play when an electron moves relative to an inertial reference frame. Thus the relative speed with which the proper inertial reference frame of the electron moves relative to the one in which the observer finds him/herself determines the applicable de Broglie wavelength that relates to an actual momentum.

Consider, for example, a sinusoidal matter-wave travelling along a single direction in space and an observer travelling with it. According to the observer the wave-intensity will not change with time in any way or manner; thus it cannot have momentum. In contrast, any observer moving relative to the proper reference frame within which the wave is stationary will attest to the fact that the wave can transfer momentum, and that this momentum is directly proportional to the speed with which the observer moves relative to the matter-wave's proper inertial reference frame. Thus Heisenberg's uncertainty relationship for a time-independent stationary wave cannot have anything to do with the wave's actual momentum as measured by an observer travelling relative to the stationary wave. All that Heisenberg's uncertainty relationship can imply is that when the wave gets "squeezed" in position space, as referenced relative to the wave's inertial reference frame, it will expand in the concomitant k-space and *vice versa*. It will now be argued that this gives an additional reason why an electron can be modelled as a wave-intensity under all circumstances. Owing to the ability of the wave intensity to "stretch and squeeze" it can manifest as an extended (delocalised) wave within a solid while "collapsing" to form a localised wave-intensity outside the solid. This is so simply because the boundary conditions (experienced by the wave) change; it has nothing to do with "implicate magic" or "wave-particle" duality. Thus the change in the wave-function when the potential energy is time-dependent (see Eq. (3)) does not necessarily describe movement of a particle, but the morphing of the wave in shape and size. It can, for example, morph relative to its proper inertial reference frame without changing the position of its centre of mass. In such a case the wave remains stationary relative to its proper reference frame. When suitable boundary conditions are encountered (changes in the applicable potential energy) it might, however, also be able to morph in such a way that its centre of mass (and thus its centre of charge) accelerates. It will then start to manifest momentum relative to its original proper inertial reference frame.

There is now, however, another serious question: why will an electron wave become localised when it enters free space? Does this mean that it forms a wave packet as had been modelled in section 4? If it does form such a wave packet, why does it do so? Furthermore, even if it does start off as a wave packet, the packet should immediately start spreading. As already pointed out in section 5 this is at variance with all experimental data collected on the free electron. If the electron is a wave, then it must become localised when it is ejected from the material in order to act as a "particle" and keep on acting as a particle while it is in free space. To do this, there must be a suitable change in the boundary conditions which the electron-wave experiences when it enters free space.

**7. Free electron localisation**
The indications are strong that a wrong assumption has been made consistently through the years when modelling the quantum-mechanical nature of a single electron moving through free space. What could this be? In contrast, it has been consistently found that the Schrödinger equation gives eminently reasonable answers when applied to all other situations where the electron is confined to, or bound to remain within a smaller region in space. The mistake that could have been made is most probably the assumption that the potential energy of a free electron is zero. This is correct when using classical mechanics; but is it also correct when quantum mechanics applies? According to Einstein's theory of relativity, mass is energy. So is this not an indication that a free stationary electron experiences potential energy? As argued in section 2, mass is a measure of the resistance of such a particle against moving from its stationary position within its proper inertial reference frame. This implies that there could be an "opposing force" coming into action when the electron "tries" to move from its stationary position; i.e. the electron is experiencing a minimum potential energy. What could be the cause of this?

As postulated in section 3 an atomic electron is a time-independent holistic localised wave that forms around the nucleus. It morphs to form such an entity because of the presence of the positive charge on the nucleus. The logic of the reasoning above indicates that a 'free' electron must also be a stationary, localised, time-independent wave-entity, or else it cannot be stationary within its proper inertial reference frame as is required for a particle with mass. It thus seems that there must be "some sort of positive charge" involved that "kicks-in" when the electron attempts to move from its stationary position. It will thus be postulated that such a charge can manifest and that this could, should and most probably must account for the mass of the electron.

Consider a solitary, stationary electron situated at the origin of its proper inertial reference frame. It is now postulated that when it attempts to move from the origin, a "virtual" positive charge appears that exerts a restoring force on the electron. It will be assumed that the space surrounding the electron is isotropic so that the same force will manifest along any radial direction. Thus if the electron moves a distance r from the origin, there will be a virtual positive charge q(r) so that the restoring force F(r) can be calculated from Coulomb's law as:

$$F(r) = -\frac{eq(r)}{4\pi\varepsilon_0 r^2} \qquad (14)$$

According to Gauss' law an electron at a radial position r should experience the same force whether the virtual charge is a point charge q(r) at the origin, or whether the virtual charge is radially distributed with a density $\rho(r)$ within a sphere of radius r centred at the origin. One can thus calculate q(r) from $\rho(r)$; i.e.

$$q(r) = \int_0^r 4\pi r^2 \rho(r) dr \qquad (15)$$

The functional dependence of $\rho(r)$ is not known. Thus a constant average value $\rho$ will be assumed to apply so that Eq. (15) becomes:

$$q(r) = \left(\frac{4}{3}\pi r^3\right)\rho \qquad (16)$$

At a suitable critical radius $R_w$ one must have that $q(R_w) = +e$, so that Eq. (16) can be written as

$$q(r) = e\left(\frac{r}{R_w}\right)^3 \qquad (17)$$

Substituting Eq. (17) into Eq. (14) leads to:

$$F(r) = -\frac{e^2 r}{4\pi\varepsilon_0 (R_w)^3} = -Kr \qquad (18)$$

One thus obtains a restoring force constant:

$$K = \frac{e^2}{4\pi\varepsilon_0 (R_w)^3} \qquad (19)$$

According to Eq. (18) the mass of an electron relates to it experiencing a harmonic potential well $V_e(r)$ within its proper, inertial reference frame, where:

$$V_e(r) = -\int_0^r Kr\,dr = \frac{1}{2}Kr^2 \qquad (20)$$

It should be noted that exactly the same potential-energy well will manifest for a positron within its proper inertial reference frame. In this case the virtual charge will be negative. Intuitively it is a very satisfactory result that the inertia of an electron (or positron) relates to being bound quantum-mechanically within a harmonic potential well; however, this has been obtained at the cost of postulating a virtual charge that appears "magically" from nowhere. Is there a mechanism that can explain the existence of such a charge? In physics it is customary to first solve simpler problems and then to extrapolate in order to find solutions to more complex ones. Many times a solution is first sought at lower dimensions. In the present case the potential well is three-dimensional. A similar problem in two dimensions, which relates to n-type semiconductors [4], will thus be considered next.

### 8. A two-dimensional analogue

In a perfect n-type semiconductor all the conduction electrons are supplied by donor centres. It is believed that in large band-gap insulators like diamond the conduction band could have a higher energy than the vacuum level outside the material. It has also been surmised that if such a material can be doped n-type it could form an ideal cold cathode. It might even be possible that the energy levels of such donor centres could also be above the vacuum level. Nonetheless, whether the energy level of the donors is above or below the vacuum level, this energy is the lowest energy a conduction electron can have within such a material.

Consider an n-type semiconductor substrate with an idealised large flat abrupt surface. The depth of the conduction band energy below the vacuum level is given by the electron affinity ($\chi$); usually measured in eV. Assume that a conduction electron gains enough energy to scale the latter barrier and move away from the substrate with a speed v. Polarisation below the surface of the substrate will generate an image charge which attracts the electron back towards the substrate. The electron decelerates and loses all its kinetic energy (some by radiating electromagnetic waves) until it, momentarily, comes to rest having only potential energy. The electron finds itself in the electric field of its image charge and will thus be at a lower potential energy than the vacuum level, which now has to be measured at infinity. The image force will then accelerate the electron back towards the substrate. This implies that the electron will be gaining kinetic energy at the expense of losing potential energy. Since the electron accelerates, it should radiate away (at least some) of the kinetic energy it is gaining. When it reaches the surface, its residual energy will be less than the vacuum level.

While moving to the surface, the electron's image charge will also move towards the surface. The image charge is a virtual charge generated by the polarisation of the conduction electrons relative to the positive donor charges within a layer just below the surface. When the image charge reaches this layer it cannot be a virtual charge anymore but has to attach itself to one of the donors. Thus the nearest distance to the surface that such a positive charge can be at, will be determined by the shallowest donor below the surface underneath the externally approaching electron. The positive charge will remain stationary at this site so that the approaching electron will experience a Coulomb potential that is similar to what an electron experiences in the radial direction around the nucleus of a hydrogen atom. This is illustrated in Fig. 1. In order to enter the substrate, the electron has to combine with this donor charge. Since the electron has radiated away energy, its energy when it reaches the surface might be below the energy of the donor level. Since the donor level is the lowest energy level available within the substrate, the electron cannot enter the substrate at all; except if it can be excited by some mechanism to reach the donor-level energy. If there is a z-axis perpendicular to the surface that goes through the positive donor charge situated at a depth s below the surface, while the electron is at a height z above the surface and displaced by a lateral distance r relative to the z-axis, the external electron will have a potential energy given by:

$$V(r, z) = -\frac{e^2}{4\pi\varepsilon_0 \sqrt{(z+s)^2 + r^2}} \qquad \text{for } z \geq 0 \qquad (21)$$

It is assumed that the coordinate r is isotropic along all lateral directions. Since r is expected to be small compared to z, the potential energy can be written as:

$$V(r, z) = V_P(z) + V_L(r, z) = -\frac{e^2}{4\pi\varepsilon_0 (z+s)} + \frac{1}{2}\left(\frac{e^2}{4\pi\varepsilon_0 (z+s)^3}\right) r^2 \qquad (22)$$

In order to describe the external electron quantum-mechanically, Schrödinger's equation has to be solved for this potential energy. This has been done elsewhere [4] and will not be repeated here; only the results relevant to the present discussion will be summarised. The wave-function can be divided into a perpendicular and a lateral component by first solving along the z-coordinate, using only the first term in Eq. (22). The possible energy levels obtained are exactly the same as for a hydrogen atom: i.e.

$$E_n = -\left(\frac{e^2}{4\pi\varepsilon_0}\right)^2 \left(\frac{m_e}{2\hbar^2 n^2}\right) \qquad \text{where n=1,2,3 etc.} \qquad (23)$$

For each concomitant perpendicular wave-function the most probable position of the electron can be calculated and then the uncertainty in the position $\Delta z$. It is then found that the allowed solutions are determined by the depth (s) of the donor charge. The deeper the donor charge (i.e. the larger s), the larger the value for the quantum number n of the lowest energy wave-function that can manifest external to the substrate. It has been found that this is determined by the Heisenberg uncertainty relationship for position and momentum. As soon as the depth s requires that the most probable position (of the electron) must be closer to the surface than the uncertainty Δz, the solution cannot manifest. This really is an astonishing result, because the surface is represented by a perfectly flat plane, as is customary when analysing interfaces in solid state electronics. Such an abstraction cannot exist. The surface of a clean semiconductor is usually reconstructed, so that there are no "dangling" bonds; i.e. all the surface orbitals are bi-electron orbitals. Thus what really should happen is that Pauli's exclusion principle prevents the external electron from becoming part of such a surface orbital. This implies that if the external electron has formed an external orbital, it can only exist if it does not have to overlap by more than a critical amount with a surface orbital. We will come back to this overlapping mechanism in section 10.

The second term in Eq. (22) has to be included next. If the most probable position along the z-direction is calculated relative to the positive charge and found to be $\langle d_n \rangle$ for the solution with quantum number n, one can replace $(z+s)$ in the second term with this most probable distance: i.e.

$$V_L(r) = \frac{1}{2}\left(\frac{e^2}{4\pi\varepsilon_0 \langle d_n \rangle^3}\right)r^2 = \frac{1}{2}K_n r^2 \tag{24}$$

Again a harmonic potential well! Although the wave function has to be two-dimensional, the Schrödinger equation along the lateral direction is in effect one-dimensional because there is no orbital momentum involved [4]. All directions are equivalent so that the two-dimensional wave is in any direction the same as for a one-dimensional wave. One thus obtains the well-known energy spectrum given by:

$$E_h = \left(h + \frac{1}{2}\right)\hbar\omega_n \tag{25}$$

where one has that:

$$\omega_n = \sqrt{\frac{K_n}{m_e}} = \sqrt{\frac{e^2}{4\pi\varepsilon_0 m_e \langle d_n \rangle^3}} \tag{26}$$

Thus for each value of n along the perpendicular direction one can have a set of solutions for h=0,1,2,3 etc. along the lateral direction. The allowed possible external energy levels are thus given by $E_{nh}$: i.e.

$$E_{nh} = E_n + E_h = -\left(\frac{e^2}{4\pi\varepsilon_0}\right)^2\left(\frac{m_e}{2\hbar^2 n^2}\right) + \left(h + \frac{1}{2}\right)\hbar\omega_n \tag{27}$$

For h=0, the wave-function has a Gaussian shape along any direction parallel to the surface; i.e. it is a zero-point wave-function. Its uncertainty in position along any direction follows as:

$$\Delta r_n = \sqrt{\frac{\hbar}{2 m_e \omega_n}} \qquad (28)$$

If $E_{nh}$ is lower than the donor energy-level, and the depth of the donor allows it, the electron can form such an external wave on the surface. To do this, the electron does not need to scale the electron-affinity barrier and lose energy by being decelerated and accelerated back (as described above); it can radiate energy by means of a "quantum jump" into the available external energy level.

When comparing the force constant given by

$$K_n = \frac{e^2}{4\pi\varepsilon_0 \langle d_n \rangle^3} \qquad (29)$$

with Eq. (19), it is clear that the force constants K and $K_n$ are exactly the same when setting $R_w = \langle d_n \rangle$. The only difference is that, in the case of the electron external to the n-type substrate, the harmonic potential energy is active in two dimensions while for the free electron it is active in three dimensions. Could the positive charge in the latter case be situated along a fourth dimension separated from the electron's charge-centre by a distance $R_w$? This possibility will now be investigated.

## 9. Separation of matter from antimatter?

By analogy it will now be reasoned that in the case of a free electron there is a positive charge situated along a fourth space dimension which is perpendicular to the three-dimensional space within which the electron manifests as a localised wave. Such a model implies that there must be a three-dimensional barrier that prevents the electron from reaching the positive charge and combining with it. What does the charge on the "other side" of the three-dimensional barrier relate to? The most logical deduction is to conclude that it relates to antimatter; i.e. to a positron. A similar model has been proposed by Boeyens but for another reason [5]. But what could cause such a barrier? "Outside" the electron-wave the fourth dimension relates to time. Since it is here assumed that the wave is a holistic entity this cannot be the case within the wave (see section 3). One can thus argue that it is for this reason why the fourth dimension is experienced as a space axis perpendicular to the isotropic three-dimensional space occupied by the holistic wave-entity (see also section 17.1). It seems intuitively logical to speculate that the three-dimensional barrier must have something to do with the existence of time; it probably relates to the curvature of space-time around an object with mass.

What is very attractive about the separation of matter and antimatter, across the three-dimensional vacuum, is that it gives a reason why our universe is mostly devoid of antimatter. This probably manifested when "our" universe was created. There is, however, a problem: the localised wave-intensity, which is assumed to be the electron, decays exponentially to infinity. So if the wave-intensity is the electron, does this not imply that the electron does not have an "outside space" around it? If this is the case, a barrier cannot form "outside" the electron.

**10. The essence of an electron-wave**

*10.1 Covalent bonding*
Consider the following experimentally verifiable situation: Two hydrogen atoms approach each other. Owing to the exponential decay in the intensities of their electron-waves (s-orbitals) these orbitals will always overlap. If the orbitals are the electrons within which their charges are distributed, does this not imply that their charge-distributions are always overlapping? It is well established that when these two atoms come "near enough" to each other they bond chemically. Why do they not do so at larger distances? Here the chemists know something that the many-body and solid-state physicists generally ignore when they do perturbation calculations.

According to chemistry, each atom has a so-called van der Waal's radius. When the atoms are far apart so that their van der Waal's radii do not overlap, one can describe the charge interactions between them (i.e. the polarisation) in classical terms. Once their van der Waal's radii overlap, chemical bonding becomes possible (in fact mandatory); i.e. it requires quantum mechanics to "kick-in". From this viewpoint, one can argue that chemical bonding occurs owing to a quantum-mechanical "short-range" force manifesting. The experimental evidence which shows that it is possible to model the charge interactions classically (i.e. as point charges) at larger distances, implies that the distributed charge of both s-orbitals does not extend further than their van der Waals' radii. One has to conclude that there exists a critical radius in all directions around the centre of charge (also corresponding to the centre of mass) defining a volume within which the essence of the electron is manifesting as distributed charge. Outside this volume the electron orbital must then be experienced as a point charge; i.e. a "point-particle" with a point charge $-e$ and a concomitant point mass. The reason why the latter situation manifests is a direct consequence of the validity of Gauss' law.

In section 8 it was pointed out that an external orbital can only exist when it does not overlap a bi-electron orbital by so much that it has to violate Pauli's exclusion principle. The same argument may now be used: the two approaching s-orbitals can manifest separately as long their van der Waals' radii do not overlap. Thus a localised wave or orbital must have (along all directions) a critical radius within which the essence of the "wave behaviour" of the holistic wave resides and which requires quantum mechanics to "kick-in" when (in any direction) the concomitant radius starts to overlap with another wave-entity's critical radius. Such a radius must be proportional to the uncertainty $\Delta r$ (calculated from the centre of charge) in the direction of overlap. We will thus write the overlap radius $R_O$ along a direction r as:

$$R_O = \beta \Delta r \tag{30}$$

The proportionality constant ($\beta$) will be called the overlap parameter.

*10.2 The Mott transition*

Another example which indicates that such a critical radius might exist is given by the Mott transition. The latter transition manifests within doped semiconductors when the dopant density reaches a critical density. Consider, for example an n-type semiconductor at very low temperatures so that the dopant electrons all reside on the donor centres. It is well known that below a critical density, the dopant-electrons are identical, individual pseudo-particles which can, under suitable circumstances, be represented by hydrogen-like s-orbitals. When the donor-density reaches the critical density, the localised s-orbitals morph into extended waves stretching over the whole volume of the semiconductor; i.e. a so-called impurity band forms. Many studies have been done on the Mott transition. A relationship, which seems to be the same for all doped semiconductors, has been found between the Bohr-radius $a_B$ of the s-orbitals and the density of dopant centres $N_D$ [6]: i.e.

$$(N_D)^{1/3} a_B \approx 0.26 \tag{31}$$

Two s-orbitals, out of a row of orbitals at this critical spacing, are shown in Fig. 2. From Eq. (31) one can derive a value for the overlap parameter as defined in Eq. (30). By setting $2R_O = (N_D)^{-1/3}$ the concomitant value found for $\beta$ is equal to $\approx 1.28$. This value is probably incorrect because the donor centres do not form a perfect cubic arrangement within the material. One expects that a percolation threshold has to be reached and that this will require a different density $N_D$ than would be the case for a perfectly periodic arrangement of the dopant centres.

*10.3 Gaussian waves*

As already discussed in section 8, external orbitals should be able to form on the surface of a suitable n-type semiconductor and that these orbitals could then have Gaussian shapes along directions parallel to the surface. It is thus probable that the density of an array of such orbitals could reach the critical overlap distance. In Fig. 3 two Gaussian orbitals out of a row of orbitals are shown when they are far from each other, and when they overlap so that just a small dimple is present between them. When they approach each other slightly more, this dimple disappears completely. This happens when the overlap parameter reaches an intriguing value; i.e.

$$\beta = \sqrt{2} \approx 1.41 \qquad (32)$$

This value does not differ very much from the value derived from the Mott-transition data. When calculating the constant in Eq. (31) using the value for β in Eq. (32), one obtains a value of 0.24; which might be the correct value that would apply when the dopants form a perfect periodic array. At this critical juncture the (pseudo) s-orbitals morph to form delocalised waves. What will the Gaussian orbitals do?

When the orbitals are single-electron, Gaussian-shaped waves, adjacent waves with opposite spins might combine to form an array of bi-electron Gaussian waves. Such a wave has zero spin and an energy equal to $E = \hbar\omega$; i.e. it is a boson. The two single-electron waves might not even need to move into each other to reach full overlap. As soon as they reach critical overlap at the overlap distance, they might already constitute a boson entity. A more interesting situation arises when all the Gaussian waves are bi-electron waves that reach overlap. Will they manifest a Mott transition to form delocalised waves over the surface of the substrate subject to Fermi-Dirac statistics? One could, alternatively, argue that each wave now manifests a boson. Will they then not rather form a Bose-Einstein condensate? One expects that this might happen under suitable circumstances for s-orbitals: for example, when helium atoms condense to form a superfluid; however, in the case of Gaussian-shaped waves a far more interesting phase transition might be able to manifest. We will analyse the latter possibility in section 16.

*10.4 Outside the critical radius*
The following question now seems to be relevant: why does the wave-intensity decay all the way to infinity if it can represent the essence of the electron by just existing up to the overlap radius $R_O$? From the analyses reported in section 8, it was found that the rest of the wave-tail outside the radius $R_O$ can manifest within the substrate material; i.e. it is the "tunnelling tail" that can penetrate an "energy barrier". It is thus a kind of "ghost tail". The part of the wave within the overlap radius cannot reside within the substrate unless the electron can chemically interact with the material constituting the energy barrier. Could the rest of the wave intensity outside the overlap radius $R_O$ have something to do with the bending of space around an entity with mass? It is an interesting thought which indicates that a simple-smooth unification of quantum mechanics and Einstein's theory of gravity might be less difficult than had been believed up to now!

**11. Visualising quantum mechanical interactions**
According to the Copenhagen interpretation one cannot visualise what is happening in the quantum world but only observe the outcomes when such interactions have taken place. Schrödinger formulated his wave equation in the hope that the latter viewpoint could be remedied. It is reported that he has said: "*I knew of [Heisenberg's] theory, of course, but I felt discouraged, not to say repelled, by the methods of transcendental algebra, which appeared difficult to me, and by the lack of visualizability*". Heisenberg stated: "*The more I think about the physical portion of Schrödinger's theory, the more repulsive I find it….What Schrödinger writes about the visualizability of his theory is probably not quite right, in other* *word's it's crap"*. After visiting Bohr in Copenhagen, Schrödinger said in despair: *"Had I known that we were not going to get rid of this damned quantum jumping, I would not have involved myself in the business.* Born rubbed salt into the wound by stating that: *"No language which lends itself to visualizability can describe the quantum jumps."* As already mentioned, these concepts have been extended and dogmatised. It is believed today that no quantum-mechanical interaction can be visualised. It will now be argued here that the interpretation of an electron as a holistic wave-entity allows one to visualise what is happening when "quantum"-interactions occur.

*11.1 Quantum jumps of an atomic electron*
How does a photon increase the energy of an orbital electron around the nucleus of an atom? According to section 7 the energy of a holistic electron-wave is its mass. Thus a lower energy electron-orbital must have a lower mass and a higher energy electron-orbital must have a larger mass. It is now postulated that when a suitable photon impinges on an atomic electron-orbital, it "coalesces" with the orbital to form another, higher-energy single holistic wave-entity. The energy of the photon adds to the mass of the electron-orbital and this requires it to "morph instantaneously" into a higher energy orbital. To fit in with older terminology this "melting together" of the photon and holistic electron-orbital will be called "entanglement". This means that before entangling, they are two separate waves, but afterwards they form a new holistic electron-orbital within which time can still not manifest; and which **is** the excited electron. According to this visualisation there are no "electron-jumps" involved but rather wave-morphs. De-excitation is thus a disentanglement of the photon energy from the excited, holistic electron-orbital which then "instantaneously" morphs back to again become a ground-state orbital.

  The word "instantaneous" is used here because it will now be postulated that wave-morphing is not limited by the speed of light. This is why it seems as if an electron jumps instantaneously from one energy level to the other. The time required is most probably only limited by Heisenberg's uncertainty relationship for energy and time; for example, when the photon adds energy $\Delta E$ to the ground-state electron-orbital of an atom, this wave must morph within a time interval $\Delta t$, as limited by Heisenberg's uncertainty relationship. The wave can only sustain the additional energy without morphing for this time-duration. This implies that the energy of any holistic wave can "flicker" with time; i.e. the wave can borrow energy for short times without changing its identity through morphing into another wave; however, sometimes the amount is large enough to effect real changes. It is proposed here that this "flickering" must be responsible for "tunnelling", and thus also for radioactive decay; i.e. by borrowing energy, the wave entity can scale an energy barrier. It should be noted that for the interpretation of quantum mechanics proposed here, a "particle" cannot "tunnel through" a barrier. This is so because the essence of the particle-wave, within the overlap radius $R_O$, cannot enter another material without chemically interacting with it. Thus "tunnelling" must always occur by borrowing energy and then scaling an energy barrier. The apparent statistical nature of quantum mechanics most probably relates completely to Heisenberg's uncertainty relationship for energy and time. It can, for example, be argued that the statistical time-behaviour observed when a collection of excited atoms de-excite also relates to this energy-time "flickering"; i.e. various "attempts" are made by an excited orbital-wave to borrow enough energy in order to scale an "activation barrier" before de-excitation can occur. The same "flickering" is probably also responsible for the anomalous g-factor of the electron, the Lamb shift, Rabi oscillations and possibly the Casimir effect. The obvious question to ask is: where does the borrowed energy come from? According to quantum field theory the vacuum has energy and the "flicker"-energy comes from "vacuum-fluctuations". We will return to this aspect in section 14.4.

A photon must also be a holistic entity "within which" time does not manifest, or else it would not be able to morph instantaneously to coalesce with an electron-wave. It should, however, be noted that in the case of a photon this absence of time does not relate to a time-independent wave function that is stationary relative to an inertial reference frame, but to the fact that the photon always moves with the speed of light relative to any inertial reference frame. When using the Lorentz transformation to calculate the time-rate "within" a photon relative to any inertial reference frame, the answer obtained is always zero; even though light waves always manifest as time-dependent fields when observed relative to any inertial reference frame.

*11.2 Interactions between atomic orbitals*

As already mentioned in section 1, the concept of "entanglement" comes from Schrödinger. According to his interpretation, all the "particles" represented by a "multi-particle" wave are entangled. In paragraph 11.1 the word "entanglement" was used to describe the situation when holistic waves "melt together" and lose their separate identities. Does this happen under all circumstances when a multi-particle wave forms? No it does not! Although the electrons around a multi-electron atom form a multi-particle wave, they have not all lost their separate identities completely. Separate orbitals manifest; except in the case where two electrons with opposite spins "coalesce" in order to form a new holistic bi-electron orbital. Only in the latter case do the two electron wave-entities entangle to form a new holistic entity. To repeat: although all the orbitals together do form a "single multi-particle wave" around the atom's nucleus, they are not superposed in such a manner that they all lose their separate holistic identities in order to form a new holistic macro-wave. In order to distinguish the latter type of superposition from entanglement, it is proposed here that it should be called "enmeshment". According to this classification, entanglement of single-electron waves can only occur between two holistic electron wave-entities when they have opposite spins. When they have parallel spins, entanglement should not be possible.

The fascinating aspect is that each atomic orbital, whether it is a single-electron orbital or a bi-electron orbital, is still interpreted as a single holistic entity within which time cannot manifest. When calculating the intensity distribution of higher energy orbitals like p-orbitals, d-orbitals etc., it is found that the intensities of these orbitals can break up into separate parts. This implies that the single holistic wave-entity can split up into separate parts within three-dimensional space when it encounters appropriate boundary conditions. How do these separate regions stay in instantaneous contact with each other? Are they connected to each other along the fourth dimension? Most probably! This aspect will be considered further in section 14.4

One can now also argue that when two hydrogen atoms bond, the electron orbitals entangle to form a bi-electron holistic entity within which the charge of -2e is redistributed within a critical volume which (in each direction) is defined by the appropriate critical radius $R_O$. As already pointed out above, the critical volume is not necessarily a sphere; it can have different values for the critical radius $R_O$ in different directions. Using this assumption, it is possible to calculate the properties of the hydrogen-molecule's covalent bond in a much simpler way than had been done in the scientific literature up to now [4].

*11.3 The photo-electric effect and Compton scattering*

The major move away from waves started when Einstein modelled the photo-electric effect as a "collision" between a photon-particle and an electron-particle. Does this not prove unequivocally that particles must be the building blocks of our universe? Consider, however, the following mechanism: an impinging photon entangles with a holistic electron-wave in the metal and thus increases its energy. If the photon energy is equal to the workfunction of the metal, the electron-wave will morph into a free electron wave with mass equal to the rest mass of the electron; i.e. it will be stationary relative to an inertial reference frame attached to the metal. If the photon has a higher energy, the electron-wave will morph into a free electron-wave with a mass larger than the rest mass of an electron. According to Einstein's special theory of relativity this extra mass manifests as kinetic energy relative to the inertial reference frame attached to the metal. The electron will thus speed away from the metal.

Thus there is no "hard-ball" collision involved; just as there are no "electron-jumps" within an atom. Both processes are the result of the photon entangling with a holistic electron-wave thus causing this electron-wave to morph into another holistic electron-wave. Similarly for Compton scattering: in this case only part of the photon's energy entangles with the holistic electron-wave in order to transfer momentum and kinetic energy. Why does one need "particles"? Waves can do it all! And what they do, can be visualised! A further interesting aspect of Compton scattering is that the photon, which should also be considered as a holistic wave-entity on its own, now splits up into two parts; one proceeding with a lower energy and the other entangling with the electron-wave.

*11.4 Diffraction of photons and electrons*

Photons and electrons can diffract. It had been Thomas Young's demonstration of double slit diffraction of light that destroyed Newton's corpuscular theory. It is the photo-electric effect that brought back the interpretation of "light particles" now called photons. How can particles diffract? Feynman called this "the only mystery in physics". When sending in light in such a way that one photon at a time impinges on the double slits, and the result is recorded as a photon at a time on a screen on the other side of the slits, then, after many photons have passed through the slits, the intensity on the screen becomes that of a diffraction pattern. Such a pattern can only form when a wave moves through both slits and "interferes" with itself on its way to the detection screen. Does each photon move through both slits? When an experiment is set up to determine whether this is possible, it is found that the photons appear on the other side after passing either through one or the other slit, not through both; however, under these experimental conditions the diffraction pattern disappears. The same arguments are valid for electrons.

According to the Copenhagen interpretation this experiment proves that we can have different outcomes depending on how we observe what is happening. If we do not look through which slit a photon moves, nature contrives a diffraction pattern. When we do look, the diffraction pattern disappears. Thus we cannot really fathom what is happening on the quantum scale; i.e. by looking we help to create a "reality" in which the photons follow definite paths like particles. If we do not look, the quantum world is able to work its "magic" unseen. So "our" universe is not just a gambling house, it is also a magician's show; things can change when the magician covers items with a cloth and then takes the cloth away.

Although it is postulated here that a photon and an electron are both holistic waves within which time cannot exist, it has been found that these entities can "go to pieces" when they encounter suitable boundary conditions. The higher-energy atomic orbitals can separate into different parts and a photon can divide in two during Compton scattering. Thus, why is it not possible that a single photon (and an electron under suitable conditions) which approaches a double slit could enmesh with the slits and move through both slits simultaneously to in this way interfere with itself on its way to the screen? While it is interfering, it stays enmeshed until another interaction occurs, or another measurement is made. When this enmeshed wave reaches a detector screen far behind the double slits, the screen makes a "position measurement" on the wave. The enmeshed photon is then "set free" and it morphs into a localised wave. Since the enmeshed photon-wave is spread out before reaching the screen, it has a wide range of positions to "choose" from. This would mean that the enmeshed wave will stay a diffracted and self-interfering wave, until it encounters the screen. If a position-detector is used in order to determine through which slit the photon came, the enmeshed wave has to localise as soon as it has passed through both slits, and thus before any interference could have occurred. No diffraction-pattern should then be produced; exactly as is found experimentally.

A recent experiment by Afshar [7] is in agreement with the explanation given here. The experimental arrangement he used is shown schematically in Fig. 4. In this experiment laser light is diffracted through two adjacent holes from where it proceeds to a lens which focuses the light from each hole to be collected by its concomitant photon detector (see Fig. 4): i.e. each detector respectively records the light coming from only one of the holes. In this experiment it was found that the images related to the impinging intensity from each hole were identical; thus showing that 50% of the light came from one hole and 50% from the other. Accordingly it has been argued that this proves that each photon came through only one of the holes. If diffraction and interference were present there should be regions on the lens where there will be no light-intensity. If the photons came randomly from the two pinholes, they should be uniformly distributed over the lens. The regions on the lens where the intensity would be zero (if diffraction occurred) were then carefully covered by wires. The images collected at the two detectors were compared to the images before these regions had been covered. Within experimental error the images were found to be the same: i.e. no photons encountered the wires on the lens. With one of the pinholes covered, the image in its respective photon detector deteriorated; in accordance with photons encountering the wires on the lens. The conclusion reached is as follows: although it is observed that the particles go through one hole at a time, diffraction is also observed simultaneously, and therefore Bohr's principle of complementarity is violated! In terms of the interpretation given here, the experimental result is, however, generated as follows: each photon does enmesh with the slits, split up and passes through both holes so that it interferes with itself on the other side. Therefore it cannot scatter from the wires, but as soon as it enters the lens, the boundary conditions change which requires it to localise again. Owing to the symmetry of the set-up, half of the photons localise to end up in one detector and the other half in the other detector.

Nonetheless, there still remains a fascinating aspect to the localisation of a single diffracting photon when it is observed by a detector or a screen. It only localises at a single point on a screen, but after many photons have localised on the screen their statistical distribution on the screen forms the diffraction pattern that relates to the intensity distribution of a single photon before it strikes the screen. There are two possibilities:

(i) The point at which the wave localises could be any point with a high enough intensity as determined by the diffracting wave before it morphs into a localised orbital. This would mean that the diffracted wave determines (on its own) what the statistical spread of collapsing should be: i.e. it acts like a probability-amplitude; albeit not quite in the manner as had been postulated by Born. This interpretation implies that when an extended photon strikes an eye it need not localise to be recorded within the eye.

(ii) Alternatively, the approaching photon could localise at the first point of encounter with the screen. One could surmise that the first wave-entity which forms part of the screen, with which the impinging diffracted wave would have to overlap by more than its overlapping radius, would be where the collapse of the wave will be recorded. Since such entities are statistically evenly spread within the screen, a large number of identically diffracted wave-entities could localise at different points in order to generate the diffraction pattern. The real reason for this statistical spread (as already surmised in section 11.1) most probably relates to Heisenberg's uncertainty relationship for energy and time; i.e. the photons, although identical, have small statistical variations in their energies (and thus shapes) and this causes them to localise at different points on the screen.

The two possibilities suggest the following experiment: change the size of the screen and see whether the intensity changes; for example, make the width of the screen the same as the width of the central diffracted line, and see whether the intensity of this line is higher than it is when using a full screen for the same number of photons. If it is not, explanation (i) applies; but if it is, explanation (ii) applies. From the viewpoint of causality, explanation (ii) would probably make more sense.

*11.5 EPR experiments*

Consider two electrons being accelerated towards each other. Their charges will repel. When they are separated by more than their critical radii, each will experience the other as a negative point charge. Solving the Schrödinger equation for either electron will cause their critical radii in the direction of approach to shrink. If their velocities are not high enough, they will scatter backwards from each other; however, if their velocities are high enough, so that their critical radii do manage to reach overlap, they could move right into each other, provided they have opposite spins. They are then able to entangle in order to form a single holistic bi-electron wave that spreads out within its proper inertial reference frame, in which its centre of mass remains stationary. If the wave is not disentangled by another interaction, it might spread for light years, and stay a single holistic entity within which time does not exist.

When a measurement is then made, which forces disentanglement to occur, the two components appear (near) instantaneously and in a correlated manner; i.e. if you choose to disentangle them with a spin measurement in a specific direction, the wave will disentangle into two sub-waves with opposite spins. An EPR experiment thus has a simple causal explanation. It is **not** two "particles" communicating faster than the speed of light, but a single holistic wave that instantaneously decomposes into two sub-waves which have to be correlated. After disentanglement, the two wave-entities are two separate single electron-waves which cannot communicate with each other faster than the speed of light. So, Einstein was right! Separate "particles" cannot communicate with each other at a speed that is faster than the speed of light. The Copenhagen interpretation is not tenable!

*11. Laser beams*

One expects that in the case of photons entanglement should be far easier because they are bosons. Many of them, having the same energy, should be able to become entangled in order to form a large macro-holistic wave-entity. I propose here that this is the essence when a laser-beam is generated. In this context one should realise that light waves with different energies should be able to enmesh; however, these separate holistic waves which form an enmeshed wave might not be separately observable when looking at the enmeshed wave with the human eye. It might require a spectrometer to separate them from each other. The latter is, of course, commensurate with all experiments done to date.

**12. Blackbody radiation**

Although Einstein cemented the quantum revolution, it was really Max Planck's analysis of blackbody radiation that gave birth to it. In order to model the blackbody spectrum, he had to assume that any light wave with a frequency ω can never have less energy than a quantum amount ΔE given by:

$$\Delta E = \hbar\omega \qquad (33)$$

This equation is at present used to argue that all light always exists of "photon-particles"; however, all it states is that the energy of light is quantised. This does not necessarily imply that light consists of "particles".

*12.1 Cavity radiation*

It is now well known that blackbody radiation can be best measured through a small hole in the side of a heated enclosed box, and that there are two ways to model the spectrum of this so-called cavity radiation; subject to the restriction given by Eq. (33):

(i) The light within the cavity is modelled as standing waves subject to Maxwell-Boltzmann statistics. Although each wave spreads right through the box (is delocalised) each harmonic wave component cannot have less energy than given by Eq. (33).

(ii) The light within the cavity is modelled as a photon gas subject to Bose-Einstein statistics. Localised "photon-particles", each with an energy given by Eq. (33) now constitute the light within the cavity.

Both approaches give identical results. There is, however, a problem that is not usually emphasised. The allowed wave frequencies are determined by the size of the cavity: the larger the cavity, the closer the frequencies (and thus energy-levels) are spaced. The fact is that these allowed frequencies can only be derived from the boundary conditions which limit the space that is occupied by delocalised standing waves. How do localised photons know which energies are allowed? After all, Bose-Einstein statistics can be applied to any distribution of energy levels. Thus to describe the radiation in the cavity by photons, one still has to first solve for the spectrum in energy levels by applying boundary conditions to waves that extend over the whole cavity. How can this be explained? One could argue as follows: it is possible for a superposed wave to be "assembled" from different basis-waves even if the basis waves are not the appropriate harmonics that will manifest under the applicable boundary conditions; for example, violin music can be constructed electronically from square basis waves (or any other shape), even though the true harmonic waves that superpose to generate the music are sinusoidal. Thus the light in the cavity most probably consists of an enmeshment of standing waves. To measure the spectrum, the light coming through the small hole must be recorded by a detector. The boundary conditions change and this requires the light wave being emitted to "de-enmesh" and morph in order to form a localised photon. Thus the measurement decomposes the enmeshed light wave within the cavity into another set of basis waves which can also represent the light in the cavity; but they are not the "real" harmonic-waves which are determined by the dimensions of the cavity. "Wave-particle duality" might thus **not** be "complementary magic", but manifests because all matter and light consist of waves which can be superposed and decomposed into different sets of basis waves other than the applicable harmonic waves.

If, as reasoned here, the light waves in the cavity are delocalised stationary waves, does it not imply that light can be stationary relative to an inertial reference frame attached to the box forming the cavity? I do not think so, because in this case the stationary waves are akin to waves on a violin string. The stationary waves are formed by light waves, all moving with light speed, which superpose. A single standing wave is in this case a superposition of a light wave (moving in a specific direction) with its component that has been reflected from the walls of the cavity (so that the latter moves in the opposite direction). Although one can still argue that it is a holistic entity, it is changing with time when viewed relative to an inertial reference frame attached to the box; i.e. it oscillates.

*12.2 Cosmic background radiation*

It has now been well-established that "our" universe is filled with black-body radiation at a temperature of ≈2.7 K. It is believed that this is the "cooled-down" version of the light that had been generated at a much higher temperature during the Big Bang. The remarkable aspect of the cosmic background radiation is its uniformity. The photons and their intensity reaching us from all directions are approximately the same. In contrast, light from far-lying galaxies is only reaching us now after it has travelled for billions of years at the speed of light; and even though these galaxies are seen as they had been billions of years ago, they look very similar to our own galaxy as well as galaxies residing closer by. It seems as if there might be some instantaneous communication across "our" universe that ensures this uniformity.

It is also well-known that "our" universe is expanding, and that this expansion is even speeding up. Thus in earlier times the universe must have been smaller. The spacing in energy levels for the cosmic background radiation must then have been larger. As the universe became larger, the spacings had to decrease. But how did the concomitant photons know that they had to adjust to these changes? Does this not imply that the cosmic background radiation has to consist of standing waves stretching right across "our" universe? But why does one measure photons when observing this radiation? In terms of the model presented in this publication this means that a light wave stretching across the whole cosmos morphs near-instantaneously into a localised photon when it encounters a detector; and this is caused by the change in boundary conditions which the detector represent. The observed uniformity of the cosmic background radiation thus relates to standing waves "filling our whole universe". As "our" universe expands, the standing light waves morphs into lower energy ones because the spacing between energy-levels becomes smaller. Therefore the cosmic background radiation cools and, owing to the ability of the waves to morph instantaneously, this happens uniformly across "our" universe. If this deduction is correct, one should, at least in principle, be able to estimate the size of "our" universe by measuring the spacing between the energy-levels of the cosmic background radiation.

**13. Bands and bonds in solids**
When theoretical solid-state physicists model the electronic properties of a covalently bonded solid material subject to the Born-Oppenheimer approximation, they find that the valence electrons, keeping the solid together, are delocalised throughout the solid and their energies lie within bands. When measuring the energies by, for example, using photo-electron spectroscopy, the energies do indeed manifest within bands. When theoretical chemists model the same valence electrons, they conclude that they pair-up to form identical bonds each having the same energy; i.e. each valence electron has the same energy. When this picture is applied to chemical interactions, it works! It confirms that under these conditions the valence electrons must all have the same energy. How can this be explained within the context proposed in this publication? Well, the solid is an enmeshed-entangled wave. Different basis waves can generate the same result. This also implies that the superposed wave can be decomposed into different basis waves by doing different measurements; i.e. by changing the boundary conditions in different ways. Thus when doing photo-electron spectroscopy, the Born-Oppenheimer approximation leads to the appropriate basis waves into which the valence electron wave can be decomposed; however, when doing chemistry the atoms play a larger role in the superposition of the valence-electron waves, and this leads to a decomposition of the enmeshed-entangled wave into identical chemical bonds.

Furthermore, when modelling a metal, the valence electron waves are all delocalised stationary-waves stretching over the whole solid. When applying an electric field via two contacts, pseudo-electrons carry a current. The charge-carriers are thus wave packets formed by the superposition of the extended waves. Do these wave packets exist when there is no electric field present which requires them to be there? It is unlikely that many such wave packets will exist when there is no electric field. If they did, it would mean that they move constantly through the solid; they should then generate electromagnetic radiation because they move relative to the other charges in the metal. This is not observed at low temperatures. Thus it seems that at low temperatures the reality is delocalised stationary waves which do not move relative to the other charges in the metal. Only when interacting with the valence electrons, by, for example, applying an electric field or heating the metal, do these waves seem to superpose and morph into wave-packets.

**14. Aspects when modelling a free electron**
The electron has always been an enigmatic entity. In order to further discuss and speculate on the model for a free electron proposed here (the basic principles have been formulated in section 7) aspects of previous analyses of the properties of an electron will also be brought into consideration and compared. This should not be considered as an attack on the previous approaches, but rather as an objective search for answers.

*14.1 The electrical self-energy field*

Even before the advent of quantum mechanics the leading physicists of the day encountered problems when analysing the electron. These problems have been summarised in *The Feynman Lectures on Physics II, chapter 2* [8]. When they tried to model the electron as a particle with a radius, inconsistencies arose. Classical electrodynamics required that the electron should be modelled as a "point-particle"; i.e. as a particle which has both its mass and charge situated at a point. Present day electron-electron scattering experiments also indicate that the electron's size could be infinitesimally small. The problem is that such a small size causes the energy of the electric field (which is calculated) around the electron to be infinitely large. This problem did not go away with the advent of quantum mechanics and is "removed" from quantum electrodynamics by means of "renormalisation". The latter procedure has become such an inherent part of all quantum field theories, that at present the "renormalisability of a theory" is accepted as proof that the theory is "realistic".

When a stationary electron with charge –e and another stationary charge q are a distance r apart, the force F exerted by the electron on the charge q is given by Coulomb's inverse square law. According to the concept of fields it is reasoned that there exists at all times an electric field E around the electron so that when placing a charge q at any position from the electron, a force $F=qE$ acts on it. According to Maxwell's equations an electric field has energy and this energy can be calculated at every point in the field from the square of the field strength $E^2$ at that point. When doing this calculation for a solitary electron, and integrating over all the points surrounding it in order to calculate the total field-energy around an electron, the energy obtained is an infinite amount. The only way to make it finite is to give the electron a substantial radius; but, as discussed by Feynman [8] even more serious problems then arise.

But does the calculation of the electric energy-field around a solitary electron not require a unique stationary reference frame to exist? The existence of other matter is, also here, ignored (similarly as had been found in section 4) and the calculation is done as if the electron is the only object existing in the universe. Where will one get a single negative charge without at least also a positive charge somewhere in the universe in order to ensure neutrality? In fact, according to electrodynamics all electric fields with energy are formed in space between separated electric charges; for example, between the plates of a charged capacitor. When, however, calculating the energy of the electric field around a solitary electron, other charges are ignored as if they are not playing any role to generate the field. This seems to be a dubious approach. After all, one can argue *vice versa* that the actual field is around the charge q, while the electron is in the field of q. In other words, how does one know that there is electric field-energy around the electron without the charge q being present? The experimental fact is that there is no other way to determine the presence of the field-energy except to use a "test" charge q. Does it then make scientific sense to assume that there is field-energy without q being part of the set-up? May it not be equally logical to assume that there is no energy at any position r around a solitary charge unless other charges are also around? If so, the derivation of the electric field-energy around a solitary electron calculates something that cannot exist.

What is ironic is that the calculation of the electric field-energy around a single electron even violates the tenets of the Copenhagen interpretation. According to the Copenhagen interpretation there are uncertainties in the position and the momentum of a point particle; so how can one have a stationary point particle around which to calculate the electric field? It might be reasoned that one can know the position of a point particle accurately but then have no knowledge of its momentum. Should one not then ask relative to which inertial reference frame can one know the position of a point particle with 100% accuracy if one does not know its momentum? One needs to also know its momentum in order to know the inertial reference frame relative to which one can specify the position of the point particle in order to calculate the energy of its "surrounding stationary field" (relative to this reference frame).

According to the approach proposed in section 7, the electrical interaction of the electron-charge is over the fourth dimension Thus the charge distribution of the electron and the charge distribution of the positron on the other side of the three-dimensional interface should cancel any electric field from manifesting within "our" three-dimensional space; just as the field is cancelled outside a charged capacitor. Thus, the model gives, in addition, a plausible reason why there cannot be an electric field around a solitary electron. Furthermore, the electron has a "soft" radius given by $R_O$. This radius shrinks when electrons are scattered from each other (see section 11.5). Thus, it is not surprising that such experiments indicate that the electron radius is extremely small. In addition, when classical electrodynamics applies, the electron is experienced by another test-charge from outside its critical radius, so that the test-charge will experience the centre of charge of the electron-wave as a point charge; consistent with all experiments to date!

There is another satisfactory aspect to the model. According to Maxwell's equations electromagnetic waves appear when charges "oscillate relative to each other". Why does an electron, far away from other matter, radiate electromagnetic waves when it is accelerated and, in addition, always produces the same radiation for the same acceleration? Could one not now argue that such accelerated motion polarizes the electron's charge relative to its positive charge along the fourth dimension, and therefore an electromagnetic field appears? After all, the positive charge supplies the electron's inertia and should thus try and drag the electron back, to in this way cause polarization and oscillatory motion relative to the electron. Furthermore, when charging a capacitor the Poynting vector is in accordance with energy flowing into the capacitor from surrounding space.

*14.2 The Dirac equation*

Dirac formulated his equation for a single free electron in an attempt to make quantum mechanics compatible with the special theory of relativity. According to the latter theory, the Hamilton-operator of an electron with mass $m_e$ and momentum p, should be written as:

$$H = \sqrt{c^2 p^2 + (m_e c^2)^2} \qquad (34)$$

The square root makes it very difficult to solve the concomitant time-independent wave equation. Dirac tackled the problem by postulating two operators $\alpha$ and $\beta$ which he used to reformulate the Hamilton operator: i.e.

$$H = c\,\boldsymbol{\alpha}\cdot\mathbf{p} + \beta m_e c^2 \qquad (35)$$

By using this Hamilton operator he could solve the wave equation to get stationary electron-wave solutions. He found that each of the resultant basis waves has to exist of two components (i.e. it is a so-called spinor); i.e. it has a direction which "explains" the spin of the electron. This has been considered as a great breakthrough because the Schrödinger equation does not supply a reason why spin should manifest (however, see section 14.5 below). In addition, he found that there are also solutions with negative energies situated at an energy-interval equal to $2m_e c^2$ below the lowest energy of the electron (which is of course its rest mass energy $m_e c^2$). This implies that a single electron cannot exist in free space; it should fall into one of the negative energy levels and disappear. To circumvent this problem, Dirac postulated that all the negative energy-levels are already filled with electrons (forming a "Dirac sea" of unobservable electrons within the vacuum) and therefore an electron with a positive energy cannot fall into them. It was then pointed out by him that if one of the negative-energy electrons can be excited by a photon with a large enough energy, it will leave a positive hole behind, which should act like a positively-charged particle. Therefore, it is widely accepted today that Dirac predicted the existence of the positron.

But also in this case, should one not ask: relative to which reference frame are these solutions not changing? Should it not be relative to a proper inertial reference frame? If this is the case, why should one use a Hamilton operator that is consistent with a particle moving at a very high speed relative to its own proper inertial reference frame? Also in this case, one can argue that the uncertainty in "momentum", as measured relative to the proper inertial reference frame, should rather give the dimensions of the wave than its momentum relative to another inertial reference frame. In addition, the operators $\alpha$ and $\beta$ seem rather contrived. The appearance of the electron spin is, however, a great plus point. It indicates that spin must have something to do with four-dimensional space-time. Another idea that caught on has been the existence of the "Dirac-sea"; i.e. that there is more to "the vacuum" than mere emptiness. This has led to the concept of "vacuum energy". It, however, lies a bit uneasy on the mind that the equation is supposedly for a single electron in space, but the result has to be interpreted as a multi-particle solution. Nonetheless, the concept of a "vacuum energy" has become an inherent part of all quantum field theories. But is the concept of "vacuum energy" not just a red herring?

*14.3 Which wave equation for a free electron?*
Is the Dirac equation the best approach when one models a free electron? My intuition says no! The Schrödinger equation is more like a regular wave equation than the Dirac equation. The possibility does, however, exist that the wave equation for a free electron could be different from both the Dirac equation and the Schrödinger equation; after all, in its time-independent format it has to calculate the rest mass of the electron if the postulates in section 7 are valid. Furthermore, the interaction between the electron and the positive charge along the fourth dimension might require a four-dimensional analogue of the Coulomb equation. This possibility does not fall within the intended scope of this publication, and will thus not be pursued any further here. Accordingly, the Schrödinger equation will be used to estimate parameters relating to the free electron in three-dimensional space. This implies that the solution in three-dimensions for the free electron relative to its proper inertial reference frame, could be similar to the lateral solution for h=0 of an external orbital on the surface of an n-type substrate. From Eqs. 2 and 25 one may thus venture to write that:

$$m_e c^2 = \frac{1}{2} \hbar \omega_e \qquad (36)$$

The frequency $\omega_e$ is determined by the distance $R_w$ along the fourth dimension. Assuming that the Schrödinger equation applies, one can write by analogy to Eq. (26) that:

$$\omega_e = \sqrt{\frac{e^2}{4\pi\varepsilon_0 m_e (R_w)^3}} \qquad (37)$$

From these equations one can solve for $R_w$ and $\omega_e$. By analogy to Eq. (28) one can then solve for $\Delta r_e$. By then using Eq. (30), an estimate for the overlap radius $R_{Of}$ for the free electron can be obtained; it comes to:

$$R_{Of} = \beta \Delta r_e \approx 2.7 \times 10^{-13} \text{ m} \qquad (38)$$

For the s-orbital of a hydrogen atom, the corresponding value $\Delta r_{se}$ calculated relative to the centre of charge of the orbital is equal to $3/2 a_B$ where $a_B$ is the Bohr radius. Accordingly, the overlap distance is given by:

$$R_{Os} = \frac{3}{2} \beta a_B \approx 1.12 \times 10^{-10} \text{ m} \qquad (39)$$

The experimentally determined value for the van der Waal's radius for the hydrogen atom is $\approx 1.2 \times 10^{-10}$ m. The correspondence with the calculated overlap radius in Eq. (39) is remarkably good.

*14.4 The fourth dimension – dark energy?*
The interaction along the fourth dimension should also add a wave- and an energy-component to the free electron. The problem is that the Schrödinger equation along this direction does not seem to lead to a reasonable solution; for example, a realistic value for $R_w$ is not obtained. As already mentioned, the solution along this direction probably requires one to use a four-dimensional analogue of Coulomb's law. In addition it probably also requires a better understanding of how general relativity and quantum mechanics meshes. Nonetheless, there is an extra energy involved. Might this be the origin of dark energy? It is an exciting thought!

This energy probably also relates to the energy that a wave in three-dimensional space can borrow subject to Heisenberg' uncertainty relationship for energy and time (see section 11.1). This could be the so-called "vacuum energy" postulated in quantum field theory. When calculating the vacuum energy according to quantum field theory, an infinite amount is obtained which has to be removed by renormalisation. If the "vacuum energy", however, relates to the fourth component of matter waves, it can never be equal to infinity, because the total amount of matter is not infinite. Furthermore, this wave component along the fourth dimension might be responsible for keeping contact between the different components of a holistic wave entity that has split up into different parts within three-dimensional space (see section 11.2).

*14.5 Electron spin*

One cannot physically visualise why the Dirac equation gives a "spinor" solution for the spin of an electron. Can the presently proposed approach (based on the principles formulated in section 7) give a reason why an electron has spin? When a magnetic field with a vector potential $\mathbf{A}$ is switched on over an electron with charge –e the momentum operator that has to be used in the Schrödinger equation becomes:

$$\mathbf{p} = -i\hbar\nabla - (-e)\mathbf{A} \qquad (40)$$

If a constant magnetic field $\mathbf{B}$ is applied, the applicable vector potential that satisfies the Coulomb gauge is the following:

$$\mathbf{A} = \frac{1}{2}\mathbf{B}\times\mathbf{r} \qquad (41)$$

Both $\mathbf{A}$ and $\mathbf{r}$ must be perpendicular to $\mathbf{B}$. If the field is applied along the z-axis then at a point with spherical coordinate r,θ,φ the vector $\mathbf{r}$ in Eq. (41) has to be perpendicular to the z-axis and its magnitude is thus given by rsinφ. The vector potential can then be written as:

$$\mathbf{A} = \frac{1}{2}|\mathbf{B}|\,r\sin\phi\,\mathbf{e}_\theta = \frac{1}{2}Br\sin\phi\,\mathbf{e}_\theta \qquad (42)$$

The unit vector $\mathbf{e}_\theta$ is the one for the θ-coordinate. The corresponding Schrödinger equation along a radial direction r then becomes:

$$-\left(\frac{\hbar^2}{2m_e}\right)\frac{d^2\psi(r)}{dr^2}+\frac{1}{2}\left(\frac{e^2B^2}{2}\sin^2\phi+K\right)r^2\psi(r)=E\psi(r) \tag{43}$$

Thus the magnetic field increases the effective force constant from K to a value K$'$; i.e:

$$K' = K + \frac{e^2B^2}{2}\sin^2\phi \tag{44}$$

This also manifests as a change in the distance $R_w$ along the fourth dimension. By using Eq. (19), it is found that $R_w$ changes to $R_w'$:

$$R_w' = \frac{R_w}{(2\pi\varepsilon_0 B^2 \sin^2\phi + 1)} \tag{45}$$

In spherical coordinates $\phi$ varies from zero to $\pi$, so that for all values of $\phi$ except $\phi=0$ and $\phi=\pi$, $R_w'$ is shorter than $R_w$. Thus the energy of the electron will be higher than the energy without the magnetic field except when $\phi=0$ (spin up) or when $\phi=\pi$ (spin down). One expects that the electron must "relax" in some manner to make either $\phi=0$ or $\phi=\pi$. This seems to imply that the direction of the fourth dimension must re-orientate in some way. If one now switches off the magnetic field and switches it on again along the same direction, re-orientation is not again required, so that the spin will remain the same; however, if the magnetic field is switched along another direction, re-orientation is again required. This is exactly how electron spin behaves, but in this case it is possible to visualise what is happening and why it is happening. One does not require a (probability) spin wave-function that defines a 50/50 probability distribution. The 50/50 behaviour occurs naturally because either spin-up or spin-down can restore the energy of the electron to its lowest level. If this requires a re-orientation of four-dimensional space within the holistic electron wave, does it not also require a change in space-time around it? This indicates that it might be possible that a magnetic field could affect gravity.

*14.6 Free positrons*

If matter and antimatter separated over a three-dimensional barrier, why can we observe positrons? Why not? If the energy supply is high enough there is no obvious reason why a situation cannot arise where an electron is situated on "the other side" of the barrier so that the positron is observed on this side. Such a state will, however, not survive for very long; as has been well documented experimentally.

*14.7 Excited electrons*

If the free electron is a ground-state holistic wave which can be calculated from a harmonic wave equation, then it is possible that an electron could have excited states. This would neatly explain the existence of the muon and the tau particles; which are in all respects like an electron but with much larger masses. Obviously, the masses of these two particles cannot be explained by the higher energy harmonic states given by Eq. (25). Since the fourth dimension also plays a role in the energy, it is likely that when an excited free electron forms, the distance along the fourth dimension $R_w$ might also have to change. If one assumes that Eqs. (36) and (37) also apply to the muon and the tau, one can calculate the concomitant changes in $R_w$ as a function of mass. The results for the electron, muon and tau, as normalised relative to the electron ($m_e=1$ and $R_w=1$) are compared in Fig. 5. A value for the proton is also included (see section 15 below).

The applicable relationship between the mass m (where m can be either $m_e$, $m_\mu$, $m_\tau$ or $m_P$) and the concomitant distances $R_w$ can be derived and is given by:

$$R_w = \left(\frac{\hbar^2 e^2}{16\pi\varepsilon_0 c^4}\right)^{1/3}\left(\frac{1}{m}\right) = 4^{-1/3}\left(\frac{\hbar}{c}\right)\frac{\alpha^{1/3}}{m} \qquad (46)$$

The fine structure constant $\alpha$ has been inserted for fun! According to this formula, an increase in mass-energy causes the distance $R_w$ to decrease inversely. Is it possible that at very high energies this distance could become so small that the particle and its anti-particle along the fourth dimension combine to form a mini black hole; i.e. the excited electron "disappears" from "our" three-dimensional space?

**15. Nucleon bonding – a single fundamental force?**

If an electron is a localised wave generated by a restoring-force manifesting owing to the presence of a positive charge across the fourth dimension, one expects that a proton could likewise be a localised, time-independent wave formed by the presence of a negative charge across the fourth dimension. When doing the same analysis as for the electron, one can also derive an overlap parameter for the proton. Using the proton's mass to determine its critical radius it is found that:

$$R_{OP} = \beta\Delta r_P \approx 1.48\times 10^{-16}\,\text{m} \qquad (47)$$

The experimentally determined radius of the proton is $\approx 10^{-15}$ m. One might now be brave and venture to argue that the radius of the proton has not been determined accurately enough by experiment; i.e. it is in reality the same as the overlap radius derived from the model, which assumes that the proton, like the electron, is a localised holistic wave-entity which forms within the electric-field of an anti-proton separated from it along the fourth dimension.

It has been concluded that chemical bonding occurs when the critical radii of atomic electron-orbitals overlap (see section 10). It thus seems attractive to conclude that nucleon bonding should similarly occur when the critical radii of two adjacent protons overlap; however, the latter requires very high energies to achieve, because when two protons approach one another, their critical radii must shrink in the direction of their approach (as explained for free electrons in section 11.5). This will, however, not happen when a proton and a neutron approach one another. This is so because the proton-charge will then experience the neutron-wave as uncharged. Is it then not possible that an electron and neutrino entangle with a proton in order to form a neutron, which can then bond quantum-mechanically with the proton in a similar way as electrons form chemical bonds. If this scenario is correct, it would imply that there are no separate "short-range" nuclear forces. The weak force relates to the entanglement of a proton with an electron-neutrino, and the strong force to a "chemical-type" entanglement of a proton and a neutron; i.e. the same quantum-mechanical mechanism might be responsible for both chemical and nucleon bonding. This further implies that that there is only one fundamental force in the universe, namely the electro-magnetic force (excluding gravity which according to Einstein is not a force at all but relates to space-time curvature). Einstein has always claimed that the only fundamental force is electromagnetic! Furthermore, the localised waves generated for a free electron and a free proton, are given by the ground state solutions of the appropriate wave equation when it is solved within the electric fields of the charges across the fourth dimension. This implies that both these particles should have excited states. Would it not be ironic if the "fundamental particles" have all along been the electron, proton, neutrino and photon, while all the other "particles" are just excited states of the electron and the proton?

It has been deduced from proton-proton scattering experiments that protons, in contrast to electrons, have "internal structure" which is at present ascribed to the presence of quarks. A free quark has not yet been generated. Much larger energies are involved for proton-proton scattering than electron-electron scattering. Might it be possible that the protons can form excited states that have been modeled by invoking quarks and gluons? The allowed excited states of the proton might thus be quite different from those for the electron and this might relate to different interactions along the fourth dimension. As already mentioned in section 10.4, to understand this probably requires a better understanding of how quantum mechanics meshes with the general theory of relativity along the lines suggested in the present publication. This does not, however, imply that all the work that has been done on quantum field theory and particle physics has been futile. All the collected data and their symmetrical interrelationships could give insight into the role quantum-mechanical interactions play over the fourth dimension.

## 16. Macro entanglement of matter

*16.1 Can matter macro-entangle?*
In section 11.6 it was surmised that a laser beam constitutes a macro-entanglement of photons. For electrons it has been deduced (in section 11.2) that entanglement can only occur for pairs that have opposite spins; however, in section 10.3 it has been argued that when bi-electron orbitals overlap they might enmesh to form a Bose-Einstein condensate; however, in the example considered it was found that bi-electron, Gaussian waves might reach a critical overlap at which they lose their individual identities; i.e. there are no separate bumps to distinguish them from one another. Could this not indicate that (at that point) these Gaussian-shaped waves become entangled, to, in this way, form a single holistic macro-entanglement; similar to photons forming a laser beam, but now consisting entirely of electrons? In fact an experimental result has been "stumbled upon" which seems to imply that this is exactly what happens.

*16.2 An experiment with electrons*
The original purpose of the experiment was to test diamond for cold-cathode action [4,9]. The diamond was doped n-type by oxygen-ion implantation. To ensure that the near-surface region had a very high density of donors, the implantation was done within a plasma-reactor at a high vacuum. The idea was to try and extract electrons by means of an anode from the doped diamond at room temperature. It has been speculated that this might be possible in the case of diamond if it has negative electron affinity (see section 8).

      The expected mechanism was as follows: When applying an electric field between the anode and a back contact to the diamond, electrons should start to flow from the diamond into the gap between the anode and diamond surface. These electrons will leave behind a positively-charged depletion layer below the surface of the diamond which attracts the electrons back towards the diamond [4,10]; i.e. an equilibrium dipole layer forms that cancels the applied field. At a high enough voltage on the anode, the gap should become filled with electrons (from the diamond's surface right up to the anode's surface). They can then act as charge carriers, and a current should then start to flow from the diamond to the anode, provided that the electrons within the conduction band of the diamond can tunnel through the depletion layer. It is for the latter reason that the dopant density has to be made very high near the surface of the diamond. The beauty is that this experiment worked first time exactly as described above; i.e. after reaching a suitable anode voltage, which depends on the distance between the anode and diamond surface (the gap-width) a current initiated and started to flow. Cold cathode action is possible! Diamond has negative electron affinity!

      When moving through the switching voltage slowly, the current was found to show instability; it increased and fell away until one moved through the critical voltage, whence it stabilised. This jumping around in the current can be ascribed to electrons entering the anode, thus causing the number of electrons in the gap to decrease so that the current falls away, followed by an increase of these electrons from the depletion layer until there are again enough of them to enter the anode again; i.e. the electrons in the gap are accelerating and decelerating all the time. A more fascinating observation was made after the device had been switched on so that a stable current was flowing. When reducing the voltage, it did not switch off. In fact, in order to reduce the current to zero, the voltage had to be reduced to zero. Furthermore, when then increasing the voltage in the negative direction, electrons started to flow from the anode to the diamond! Since the anode was a metal without any negative electron affinity, this was quite an astonishing result. In

addition one could switch off the equipment for weeks. When subsequently switching it on, it still conducted current without having to first increase to a critical voltage. One could even let air into the system without changing the current. This could be done for voltages that, without the conducting phase, would cause a plasma discharge which damaged the diamond and the anode. Intensive experiments were lodged to see whether this was not caused by contamination; however, it had to be concluded that just before the phase forms, the device behaved and switched exactly as expected for electrons; so whatever was happening must relate to the presence of electrons within the gap.

The device was then again analysed theoretically. It was realised that even after the current started to flow, there will still be an electric field over the depletion layer below the diamond's surface which cannot be balanced by a further increase in the width of the layer of electrons outside the surface; i.e. the width cannot expand further than the surface of the anode. Thus the depletion layer will try to increase the width of the electron-layer by keeping on injecting electrons into the gap between the diamond and the anode. Instead of achieving an **increase** in the field, the density of charge-carriers within the gap will increase; more electrons are thus available to transport the current so that it causes a concomitant **decrease** in the electric field within the gap. To reach an equilibrium current, as required by the second law of thermodynamics, the depletion layer must stop adding electrons. This will only happen when the electric field over the depletion layer and the gap becomes identically zero. This implies that an equilibrium current, as experimentally measured, can only flow through the gap if there is no field between the diamond's surface and the anode. This is the definition of superconduction! Thus in order for the experimental result to manifest, as it did and still does, the electrons within the gap must in some way form a superconducting phase.

An independent evaluation of the above experiment has recently been initiated at the University of KwaZulu-Natal [11]. This work has so far been in accord with the previous findings and has also generated new insights. The equipment for the latter experiments has been constructed in such a way that the gap between the diamond surface and the anode can be observed by means of a microscope. It has been found that (at the switching voltage) during the time that the current was jumping around, light sparks emanated from the gap; just as one would expect for electrons accelerating and decelerating. Once above the critical voltage a perfect black cylinder with a diameter of ≈1 μm immediately appeared between the anode and the surface of the diamond (see Fig. 6 for a schematic representation). The colour is exactly what one expects for a superconducting phase; light (electromagnetic radiation) cannot enter or move through it. When sending large currents from the diamond to the anode, which would cause any other material of similar thickness to glow white hot or melt, the phase still stayed black. In fact, a current could be sent from the diamond to the anode so that the anode (made of steel) started to melt. When opening the gap rapidly it is possible to break the cylinder. When a piece remains on the anode or/and the diamond's surface, the cylinder is slowly sucked away into the material it is in contact with. If on the other hand the voltage is increased, the cylinder finally re-establishes contact and the current starts to flow again. Further experiments are in progress [11].

It is interesting to note that the electron phase forms between positive charges on the anode and the cathode. Thus, it is reminiscent of a giant chemical bond; just like two electrons between two protons form a covalent bond one now has many electrons forming a single entity between the positive charges on the anode and cathode surface. The only way that this can be explained is to conclude that the electrons have entangled to form a single holistic entity. How? As already argued above, the depletion layer will keep on adding electrons to the gap as long as there is an electric field within the gap. As the electron density increases, the charge carriers move slower and slower; in fact they approach zero velocity. Each electron experiences an average positive charge, less than $+e$, which is induced jointly by the positive charges on the cathode and anode and the other surrounding electrons, and thus starts to morph into a Gaussian-shaped wave by "vibrating" through the induced virtual positive charge. With a further increase in electron density these Gaussian waves finally reach overlap so that bi-electron Gaussian waves form, which, in turn, finally reach overlap to entangle with the formation of a single holistic macro-wave within which time does not manifest.

There are thus no separate charge-carriers anymore; so how does charge-transport manifest? The conclusion one has to reach is that a current cannot flow "through" such a phase because this requires time to manifest within the wave-entity. Charge-transport has to occur by "teleportation". When electrons are injected at the cathode they entangle with the wave and thus change the wave's energy. According to Heisenberg's uncertainty relationship for energy and time, this change in energy can only be tolerated for a limited time. Accordingly, electrons disentangle at the anode and flow into the anode. Faster than light signal transfer **is** thus possible just as in the case of an EPR experiment.

*16.3 An experiment with buckey balls*

It has been reported in the literature that buckey balls can diffract at low enough temperatures [12]; even though such a structure is known to consist of many carbon atoms. When the temperature is increased, the diffraction pattern becomes "washed out" and finally disappears. Furthermore, emission of photons from the buckey balls causes the diffraction pattern to disappear. This also happens when the vacuum deteriorates so that the buckey balls start to regularly encounter gas molecules. According to the discussion in section 11.4, this implies that a buckey ball must be able to enmesh with the diffraction slits and move through both of them, and this is only possible for a holistic wave within which time does not manifest; i.e. the buckey ball must for some reason be able to become an entangled single entity when it is cooled to low temperatures. When the temperature increases, the buckey ball changes from being an entangled entity to become an enmeshed entity; this is required for it to be able to emit photons. Therefore it cannot enmesh with the slits as a single holistic entity in order to diffract.

When the vacuum pressure increases, the temperature does not, so that the buckey balls should not change from an entangled to an enmeshed entity; however, they now "encounter" the gas atoms, and each encounter localises the buckey ball. Thus after a buckey ball passed through both slits simultaneously, it will be localised on the other side of the slits before its component-parts which moved through the two slits can interfere to form a diffraction pattern.

*16.4 Dark matter?*

If buckey balls can become entangled, can this not also occur for much larger masses of neutrally-charged materials? Could this not be the origin of dark matter in the universe? Such matter would be totally inert and unobservable because it is unlikely that it will interact with baryonic matter or light in a way that will make it "visible". It should primarily manifest its presence through gravity. This seems to be a good description of the properties of the dark matter which is believed to be present in "our" universe. If the dark matter in "our" universe does consists of very large volumes of entangled matter, might it not supply pathways to the stars? If one could get a spaceship to entangle with such an entity it might be teleported over distances of light years; similar to electrons being teleported from the cathode to the anode through a holistic superconducting phase (see section 16.2).

**17. Creation of the universe – bending space?**

*17.1 Four-dimensional space, time, temperature and entropy*

From Einstein's theories of relativity it is known that "our" universe exists against the backdrop of curved four-dimensional space-time. In this publication it is proposed that holistic wave entities exist within which time does not manifest; so that the entity is in immediate contact with itself over the whole volume it occupies in three-dimensional space. The proposed model for the electron in section 7, leads to a solution where a four-dimensional Euclidean space manifests "within" the electron; i.e. the fourth dimension is perpendicular to three-dimensional space. Does this not indicate that four-dimensional space-time cannot be Euclidean; i.e. for time to manifest the fourth axis cannot be perpendicular to three-dimensional space?

The latter deduction makes perfect mathematical sense. In a Euclidean space the coordinates along the axes are linearly independent; i.e. differentiating any one of them with another gives a zero result. Thus, if time exists along an axis which is perpendicular to three-dimensional space, one will not be able to differentiate any space coordinate with time in order to calculate a velocity; i.e. kinetics will not be possible! If kinetics is not possible within four-dimensional Euclidean space, it implies that such a space can also not manifest either temperature or entropy; i.e. it manifests "perfect order" and immediate connectivity.

At the event horizon of a black hole time stops. A black hole is modelled as a singularity in space-time which is "cloaked" by its event horizon. Is it not maybe possible that this singularity is a "virtual" entity similar to an image charge? Maybe the event horizon is a direct interface between "our" universe and the "surrounding", timeless four-dimensional Euclidean space. Could one not argue that there is no barrier along the fourth dimension at the event horizon; and therefore matter and light can move from "our" universe to again "coalesce" with the timeless, zero-entropy, four-dimensional Euclidean space that exists as "nothingness around our universe"?

*17.2 Alpha and omega*

Every space has a reciprocal space. When the space is perfectly isotropic and Euclidean it is its own reciprocal space. As already asked above: could this not imply that "before creation" the universe was an infinite four-dimensional Euclidean space? At creation, or the so-called Big Bang, a region of the four-dimensional Euclidean space became "bent", so that our space and its reciprocal space started to separate. Matter and antimatter separated across a three-dimensional barrier, which we now experience as our three-dimensional vacuum. If all this matter initially consisted of entangled dark matter, this matter might have near-instantaneously morphed to cause rapid inflation of "our" universe from a small size to a stupendous size; after which baryonic matter started to "precipitate" out. This might be the reason why most matter is still in the form of dark matter. Furthermore, this could be the reason why matter precipitated out in the same manner even when separated by light years. "Time" does not exist "within" the dark matter from which normal matter precipitated; i.e. all baryonic matter is the "offspring" of the same "mother-matter" which is, or was, everywhere in instantaneous contact with "herself". In addition, the formation of space-time, owing to the bending of the time-axis, made the passage of time and the existence of light to manifest. This also enabled the baryonic matter, which have "precipitated" out, to experience kinetic change. This would imply that "during creation" a high entropy state "jumped" into existence from zero entropy surroundings. So are we living in a cosmos that is evolving from high entropy to even higher entropy? This seems unlikely. Maybe our cosmos is expanding in order to "un-curve" so that at the end of time it will again "completely" merge with its infinite, timeless surroundings. This would imply that the arrow of time is determined by a decrease of the total entropy of "our" cosmos. Maybe this is why order crystallises out in the form of living beings and plants? It is interesting to speculate further that if our cosmos is just a localised region within an infinite four-dimensional Euclidean space, then there might be many more such regions. So maybe our cosmos is not as unique as we want to believe.

**18. Summary and conclusion**

In summary, the following postulates and conclusions emerged:

(i) The building blocks of matter might be holistic fields (waves) within which time does not manifest.

(ii) The essence of each holistic matter-wave is contained within a volume, the size of which is defined in any direction by a critical radius, where this radius is equal to the "uncertainty" in position relative to the centre of mass of the wave multiplied with an overlap parameter ($\beta$); the latter seems to have a universal value equal to the square root of two.

(iii) The mass of a holistic matter-wave is equal to its quantum-mechanical energy calculated (from a time-independent wave equation) relative to its proper inertial reference frame.

(iv)   A holistic matter-wave can near-instantaneously change in shape and size (morph) when the boundary conditions change or when it coalesces (entangles) with a photon in order to form another, higher-energy, holistic matter-wave entity. Similarly, the morphing is also near-instantaneous when it disentangles into sub-waves.

(v)   Holistic matter-waves can superpose in two ways; (a) they can "enmesh" without losing their holistic individual identities or (b) they can "entangle" to form another holistic matter-wave within which time also does not manifest.

(vi)   Light waves can also enmesh or entangle; as well as morph from being vast extended holistic standing-wave entities to form localised photons when the boundary conditions change.

It has been argued here that these concepts lead to a causal interpretation of quantum mechanics which can be visualised. Diffraction and EPR phenomena are explained as outcomes which should be expected to manifest when waves interact. The photo-electric effect and Compton scattering have been modelled in terms of matter-waves entangling with light-waves. It has been deduced that chemical bonding and nucleon bonding most probably relate to the same quantum-mechanical mechanism; i.e. excluding gravity there is only one fundamental force in nature namely the electromagnetic force. Indications, of how quantum mechanics and general relativity might relate to each other, emerged. It has been deduced and proposed that "dark energy" most probably relates to the energy of a wave component (of matter-waves) along the fourth dimension, while "dark matter" could be large macro-entangled matter waves within which time does not manifest. A possible explanation for the absence of antimatter "within our universe" emerged. A plausible reason has been proposed for the uniformity of cosmic background radiation.

**Acknowledgements**

I would like to thank Terry Doyle for allowing me to quote some of his unpublished results. Over the last three years I have had numerous discussions with my friend and colleague Roger Wedlake. I am very grateful for his support and enthusiasm. These discussions and especially his incisive questions have been extremely useful.

**Figure captions**

Fig. 1: The formation of an external electron-orbital (or wave) on the surface of an n-type semiconductor substrate surface.

Fig. 2: Overlapping of hydrogen-like electron s-waves: the two waves form part of an array of such waves and they are separated by the critical distance at which the Mott-transition manifests.

Fig. 3: Overlapping of Gaussian-shaped bi-electron waves: the two waves form part of an array of such waves. The dashed curves show them when they are far away from each other. The solid curve shows them when they are near the critical overlap distance that corresponds to the Mott transition. When they reach the Mott critical distance, the dimple between them disappears.

Fig. 4: Afshar's experiment: This experiment demonstrates that both wave and particle behaviour can manifest in the same experiment. This violates Bohr's principle of complementarity.

Fig. 5: A possible relationship between the distance $R_w$ over the fourth dimension and the mass of a particle. The data has been normalised relative to the electron so that the mass $m_e$ of the electron as well as its concomitant distance $R_w$ are both equal to unity.

Fig. 6: The shape and form of an entangled superconducting phase, consisting entirely of electrons, which form between the surface of an n-type diamond and an anode after a suitable electron-extraction potential has been applied.

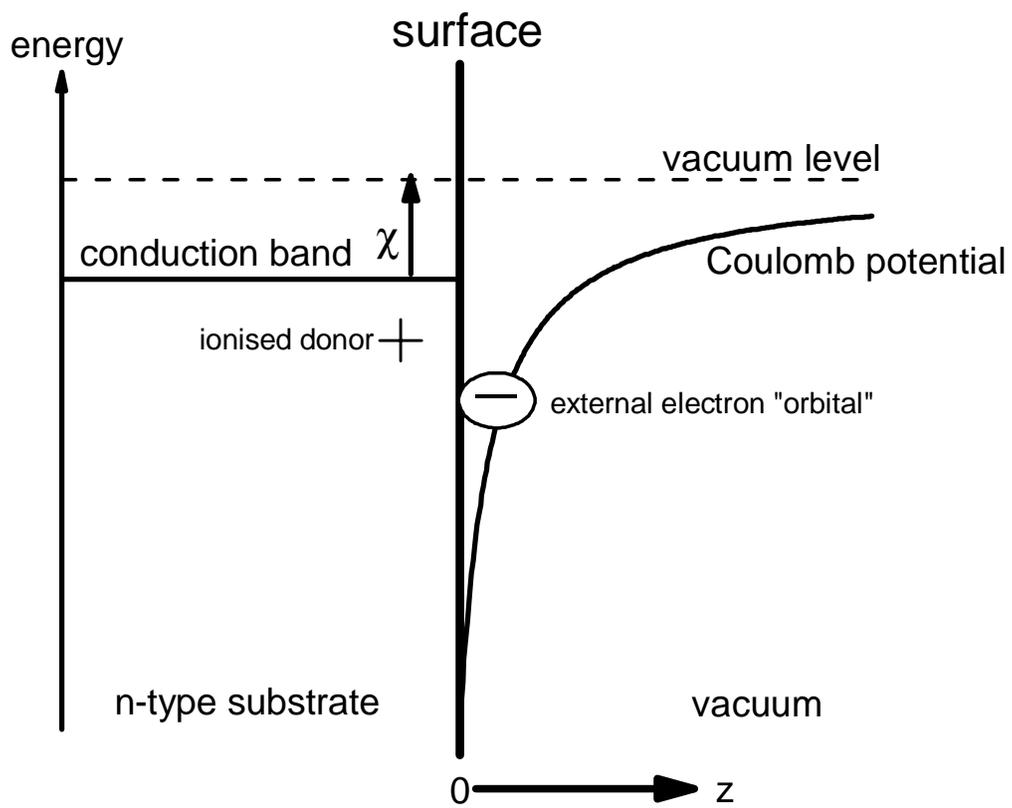

Figure 1

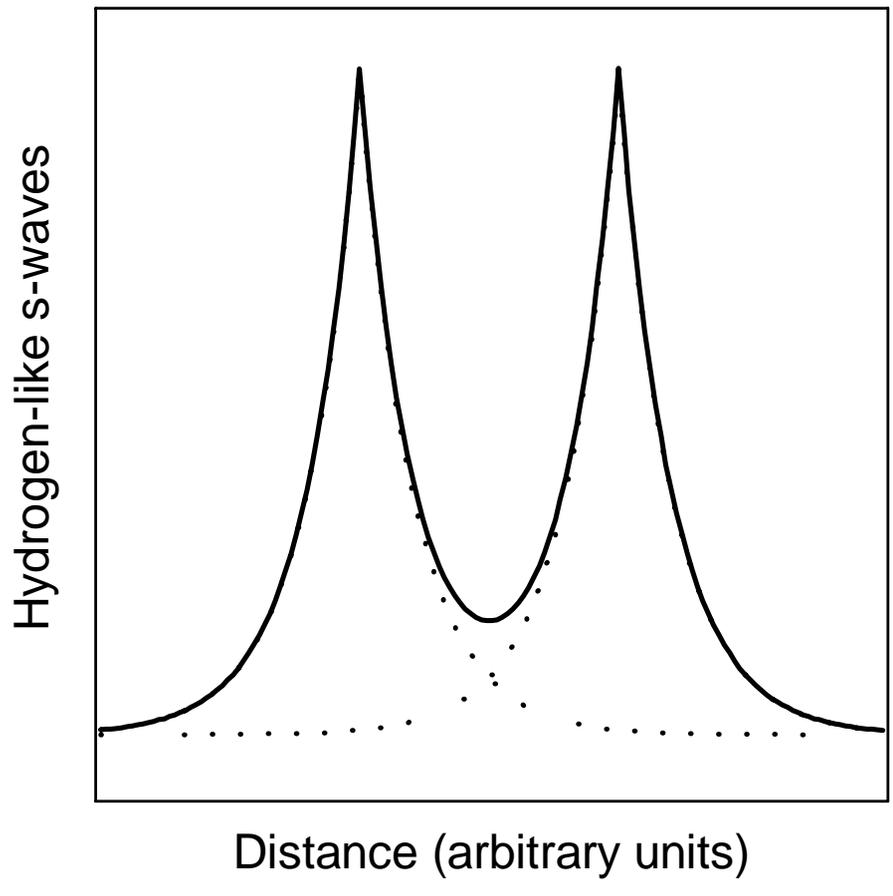

Figure 2

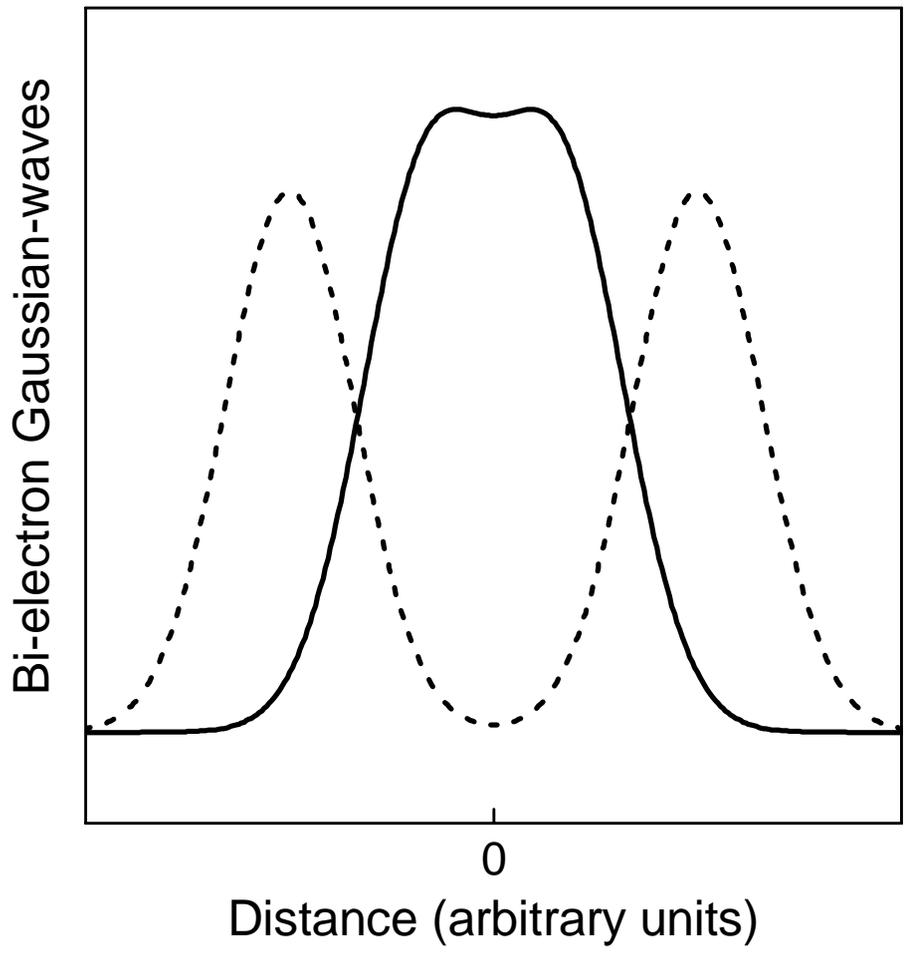

Figure 3

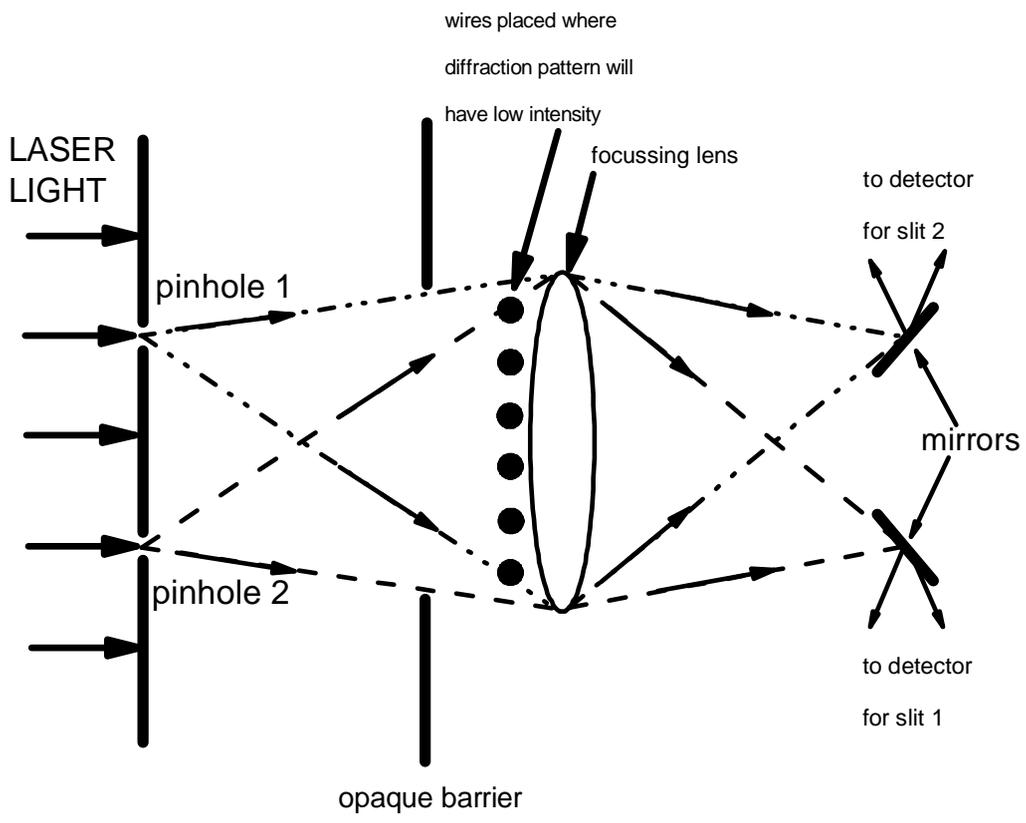

Figure 4

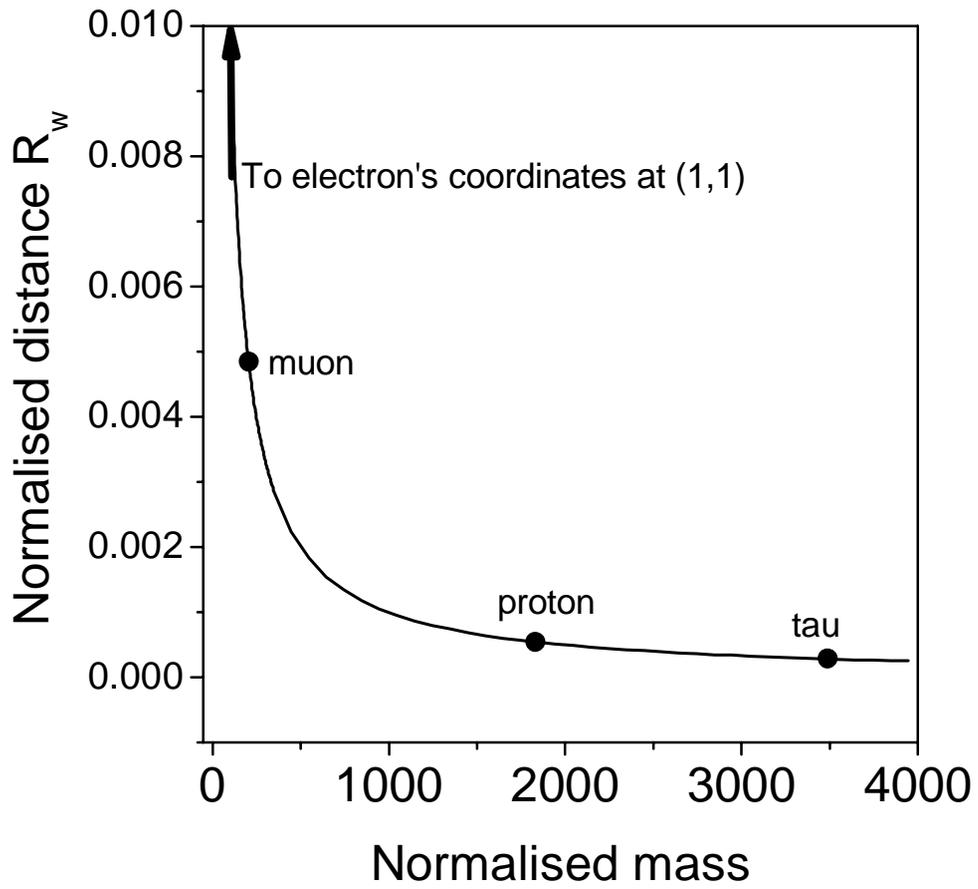

Figure 5

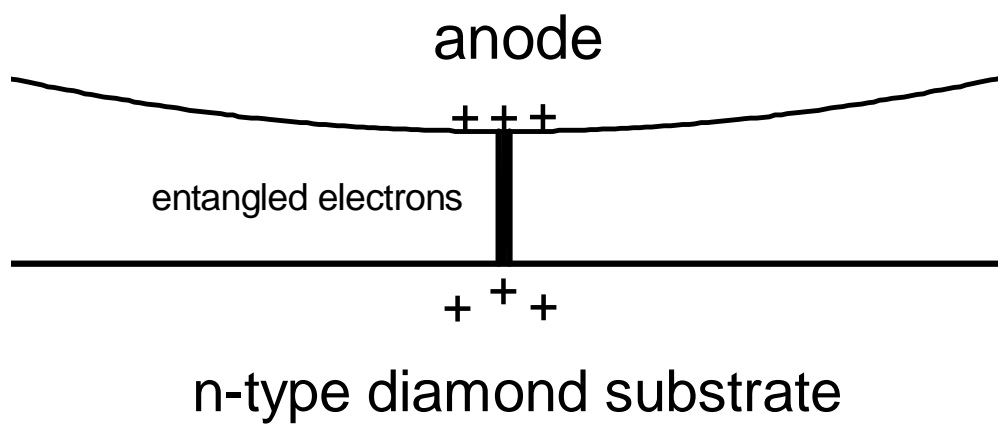

Figure 6